\documentclass[%
 reprint,
superscriptaddress,
 amsmath,amssymb,
 aps,
prc,
]{revtex4-2}
\usepackage{graphicx}
\usepackage{dcolumn}

\usepackage{bm}
\usepackage{color}

\usepackage{multirow}
\usepackage{mathtools}
\usepackage[utf8]{inputenc}

\begin{document}

 \preprint{APS/123-QED}

\title{Observational constraint from the heaviest pulsar  PSR J0952-0607 on the equation of state of dense matter in relativistic mean field model}
 \author{Raj Kumar}
  \email{raj.phy@gmail.com}%
\affiliation{Department of Physics, Himachal Pradesh University, Shimla-171005, India}

\author{Mukul Kumar}
\affiliation{Department of Physics, Himachal Pradesh University, Shimla-171005, India}
\author{Virender Thakur}
 \email{virenthakur2154@gmail.com}
 \affiliation{Department of Physics, Himachal Pradesh University, Shimla-171005, India}
\author{Sunil Kumar}
\affiliation{Department of Physics, Himachal Pradesh University, Shimla-171005, India}
\author{Pankaj Kumar}
\affiliation{Department of Applied Sciences, CGC College of Engineering, Landran, Mohali - 140307, India}
\author{Anuj Sharma}
\affiliation{Department of Physics, Himachal Pradesh University, Shimla-171005, India}
\author{B.K. Agrawal}
\affiliation{Saha Institute of Nuclear Physics, 1/AF Bidhannagar, Kolkata 700064, India}
\author{Shashi K. Dhiman}
 \email{shashi.dhiman@gmail.com}
\affiliation{Department of Physics, Himachal Pradesh University, Shimla-171005, India}
\affiliation{School of Applied Sciences, Himachal Pradesh Technical University, Hamirpur-177001, India}

\begin{abstract}

In the present work, we constrain the equation of state of dense matter in the context of heaviest observed neutron star mass M$_{max}$ = 2.35$\pm 0.17$ M$_{\odot}$ for the  black widow pulsar PSR J0952-0607. We propose  three  interactions for the relativistic mean field model which include
different combinations of  non-linear, self and cross-couplings among isoscalar-scalar $\sigma$, isoscalar-vector $\omega$ and isovector-vector $\rho$ meson fields up to the quartic order. These interactions are in harmony with the finite nuclei and bulk nuclear matter properties. The equations of state  computed by using newly generated interactions for the  $\beta$-equilibrated nucleonic matter satisfy the heaviest observed neutron star mass M$_{max}$ = 2.35$\pm 0.17$ M$_{\odot}$ for the  black widow pulsar PSR J0952-0607. The results for the radius ($R_{1.4}$) and  dimensionless tidal deformability (${\Lambda_{1.4}}$) corresponding to the canonical mass are also presented and agree well with the GW170817 event and astrophysical observations. The radius of  $2.08M_{\odot}$ neutron star mass is predicted to be in the range $R_{2.08}$ = 12.98 -13.09 Km which also satisfy the NICER observations by  Miller et al. (2021) and Riley et al.(2021). A covariance analysis is also performed to assess the theoretical uncertainties  of model parameters and to determine their correlations with nuclear matter observables.
\end{abstract}
\maketitle


\section{INTRODUCTION}
Neutron stars are stellar objects made of highly dense asymmetric matter and have extreme properties. The dense core of the neutron star enables us to study nuclear matter beyond saturation density. The  composition of the matter at such high density is not known exactly to the date, but the thermodynamic state of the matter is theoreticized by the equation of state (EoS). In recent years,  many advances in astrophysical experiments have probed  new constraints on EoS by studying properties like mass, radius and tidal deformability of neutron stars. Constraints from terrestrial experiments have been obtained by studying matter at supra-saturation density in heavy ion collisions and determining the neutron skin thickness. The possibility to detect gravitational waves from merging binary systems by the LIGO and VIRGO collaborations \cite{Abbott2018,Abbott2019} and   NICER measurements \cite{Miller2021,Riley2021} on mass radius have major contributions to probe the behavior of EoS from low to the high-density regime.
 The neutron stars are  highly dense asymmetric nuclear systems having a central density about 5-6 times the nuclear saturation density. In recent years, new measurements of masses from radio pulsars timing \cite{Demorest2010, Antoniadis2013,Cromartie2020}, tidal deformabilities from gravitational wave analyses \cite{Abbott2017,Abbott2018} and radii from X-ray pulse profiling \cite{Miller2019,Riley2019,Miller2021,Riley2021} have attracted a great deal of attention, as these measurements have started to clarify about the possible existence of the novel state of matter in the dense inner core of heaviest neutron stars and EoS of dense matter \cite{Essick2021a, Li2021a, Drischler2021}.
  The nuclear theory studies \cite{Haensel2007,Lattimer2014,Baym2018} are mainly focusing on understanding the dense matter in a neutron star. The constraints on EoS at high density are imposed with currently available lower bound  on  neutron star's maximum mass and radius \cite{Hebeler2010,Hebeler2013,Lattimer2012}. The precise measurement of masses of millisecond pulsars such as PSR J1614-2230 \cite{Demorest2010}, PSR J0348+0432 \cite{Antoniadis2013} show that the maximum mass of the NS  should be around 2 M$\odot$.
The recent observations with LIGO and Virgo of GW170817 event \cite{Abbott2018,Abbott2019} of Binary Neutron Stars merger
and the discovery of neutron star  with masses around 2$M_\odot$ \cite{Demorest2010,Antoniadis2013,Arzoumanian2018,Miller2019,Riley2019,Raaijmakers2019} have intensified the interest in these fascinating objects. The analysis of GW170817 has demonstrated the potential
of gravitational wave (GW) observations to yield new information related to the limits on neutron star  tidal deformability.\\
The PSRJ1748-2446ad discovered by Hessels et al., \cite{Hessels2006} is the fastest spinning pulsar having frequency  716 Hz ($P_{s}$ = 1.3959 ms) and mass $\le$ 2 M$_{\odot}$.
But, Pulsar PSR J0952-0607 was discovered by Bassa et al.(2017) \cite{Bassa2017} with a spin period of $P_{s}$= 1.41 ms, making it the fastest and heaviest (M$_{max}$= 2.35$\pm$0.17 M$_{\odot}$) known galactic neutron star in the disk of the milky way. It is a black widow pulsar with a low mass substellar companion being irradiated and evaporated by the pulsar luminosity. The Pulsar J0952-0607 is a particularly attractive candidate for further investigation, as the neutron star mass is indeed the largest well-measured value (M$_{max}$= 2.35$\pm$0.17 M$_{\odot}$) to date. This heaviest observed neutron star for the black widow pulsar  should have deep implications on dense matter EoS. Also with  a central value of 2.35 M$_{\odot}$, PSR J0952-0607 provides the most severe constraints on dense matter EoS \cite{Romani2022}.\\
The recent parity-violating electron scattering experiments on $^{48}$Ca (CREX) \cite{Adhikari2022} and $^{208}$Pb (PREX-II) \cite{Adhikari2021} are of considerable interest. The PREX I+II combined yield  neutron skin thickness for $^{208}{Pb}$ as $\Delta r_{np}$ = 0.283 $\pm$ 0.071 fm which implies [20] 68\% confidence ranges of of symmetry energy J  = 38.29$\pm$4.66 MeV and slope of symmetry energy  L = 109.56$\pm$36.41 MeV. Both values, and the measured  value of neutron skin thickness itself, are considerably larger than from expectations from neutron matter and value of nuclear binding energies, including the previous measurements, although overlapping with them at about the 90\% confidence level. This indicates a tension with the current understanding of the EoS.
In contrast, the measurement of the neutron skin of $^{48}$Ca using the same technique \cite{Adhikari2022} is somewhat smaller than the average of earlier experimental measurements and expectations from nuclear binding energies and neutron matter theory.
 A Bayesian analysis of the PREX and CREX results is performed in Ref. \cite{Zhang2022}.
 They found that the two experimental results are incompatible with each other at 68 percent confidence level, but compatible at 90 percent confidence level. Combining the data, they inferred J = 30.2$^{+4.1}_{-3.0}$ MeV and
 L = 15.3$^{+46.8}_{-41.5}$ MeV at 90 percent confidence level.
 They find the combined results predict $\Delta r_{np}$ for $^{48}$Ca close to the CREX result, but predict $\Delta r_{np}$ for $^{208}$Pb considerably smaller than the PREX result. A combined analysis is also performed in \cite{Reinhard2022a} and conclude that a simultaneous accurate description of the skins of $^{48}$Ca and $^{208}$Pb cannot be achieved with their models that accommodate mass, charge radii and experimental dipole polarizabilities. The two experiments CREX and PREX-II separately predict incompatible ranges of L,  L = -5$\pm$40 MeV and L = 121$\pm$ 47 MeV, respectively, but accepting both measurements to be equally valid suggests J = 32$\pm$2 MeV and L = 50$\pm$12 MeV \cite{Lattimer2023}. The large value of  $\Delta r_{np}$ = 0.283 $\pm$ 0.071 fm suggests a very stiff EoS and large value of L around saturation density that generally gives rise to a large value of neutron star radius and the tidal deformability \cite{Reed2021}. The upper limit on $\Lambda_{1.4}$ $\leq$ 580 for GW170817 requires softer EoS and hence softer symmetry energy coefficient \cite{Abbott2018}. The heaviest neutron star 2.14$^{+0.10}_{-0.09}$ M$_{\odot}$ of PSRJ0740+6620 \cite{Cromartie2020} also strongly limits the symmetry energy under the constraint on the EoS of symmetric nuclear matter (SNM) from flow data in heavy ion collisions \cite{Danielewicz2002} which is relatively soft and strongly limits the neutron star maximum mass. The fastest and  heaviest observed  galactic neutron star $M_{max}= 2.35\pm 0.17 M_\odot$ for the black-widow pulsar PSR J0952-0607 \cite{Romani2022} may put  stringent or severe constraints on the symmetry energy at  high densities and on  EoS of dense matter. The heaviest observed neutron star mass for the black widow pulsar  demands a stiff EoS and a stiff slope of the symmetry energy coefficient. \\
The motivation of the present work is to generate  new parametrizations for the RMF model that can be  used to constrain the EoSs in light of heaviest observed neutron star M$_{max}$ = 2.35 $\pm$0.17 M$_{\odot}$ for the black widow pulsar PSR J0952-0607. The RMF models used in the present work include different combinations of  non-linear, self and cross-couplings among isoscalar-scalar $\sigma$, isoscalar-vector $\omega_{\mu}$ and isovector-vector $\rho_{\mu}$ meson fields for isoscalar and isovector sectors. These parametrizations are to be  generated in such a  way so that they can  accomodate the  properties of   neutron stars within the astrophysical observations  without compromising the finite nuclei and bulk nuclear matter  properties.  The $\omega$ meson self-interaction  term  $\zeta$ plays an important role in determining  the soft and stiff behavior of EoS at high densities without affecting the bulk nuclear matter properties at high density.  The  neutron star mass decreases with the increase in value of coupling $\zeta$ \cite{Dhiman2007,Raj2006,Horowitz2001, Muller1996,Pradhan2023}. So in order to maintain compatibility with a heaviest neutron star mass, this self-interaction term is either not incorporated  or a very small value is taken in many recent studies which employ the RMF models \cite{Fattoyev2020,Virender2022a}. We also generate a parameter set in accordance with the naturalness behavior as imposed by the effective field theory \cite{Furnstahl1997}.\\
The paper is organized as follows, in section \ref{tm}, a brief outline of the RMF Lagrangian, equations of motion and  EoS for neutron stars is provided. In section \ref{parameter}, the procedure for optimization  of the model parameters is discussed. Numerical results and detailed discussions concerning features of finite nuclei, bulk nuclear matter and neutron star matter are presented in section \ref{results}. Finally, we give a summary in section \ref{summary}.
\section{THEORETICAL MODEL}\label{tm}
The effective lagrangian density for the RMF model generally describes the interaction of the baryons via the exchange of $\sigma$, $\omega$, and $\rho$ mesons up to the quartic order. The Lagrangian  density\cite{Dhiman2007,Virender2022a,Raj2006} is given by
\begin{eqnarray}
\label{eq:lbm}
{\cal L} &=& \sum_{B} \overline{\Psi}_{B}[i\gamma^{\mu}\partial_{\mu}-
(M_{B}-g_{\sigma B} \sigma)-(g_{\omega B}\gamma^{\mu} \omega_{\mu}\nonumber\\&+&
\frac{1}{2}g_{\mathbf{\rho}B}\gamma^{\mu}\tau_{B}.\mathbf{\rho}_{\mu})]\Psi_{B}
+ \frac{1}{2}(\partial_{\mu}\sigma\partial^{\mu}\sigma-m_{\sigma}^2\sigma^2)\nonumber\\  &-&
\frac{\overline{\kappa}}{3!}
g_{\sigma N}^3\sigma^3-\frac{\overline{\lambda}}{4!}g_{\sigma N}^4\sigma^4  - \frac{1}{4}\omega_{\mu\nu}\omega^{\mu\nu}
+ \frac{1}{2}m_{\omega}^2\omega_{\mu}\omega^{\mu}\nonumber\\&+& \frac{1}{4!}\zeta g_{\omega N}^{4}(\omega_{\mu}\omega^{\mu})^{2}-\frac{1}{4}\mathbf{\rho}_{\mu\nu}\mathbf{\rho}^{\mu\nu}+\frac{1}{2}m_{\rho}^2\mathbf{\rho}_{\mu}\mathbf{\rho}^{\mu}\nonumber\\
&+&\frac{1}{4!}\xi g_{\rho N}^{4}(\mathbf{\rho}_{\mu}\mathbf{\rho}^{\mu})^{2}+
 g_{\sigma N}g_{\omega N}^2\sigma\omega_{\mu}\omega^{\mu} \left(a_{1}+\frac{1}{2}a_{2}g_{\sigma N}\sigma\right)\nonumber\\
&+&g_{\sigma N}g_{\rho
N}^{2}\sigma\rho_{\mu}\rho^{\mu} \left(b_{1}+\frac{1}{2}b_{2}g_{\sigma N}
\sigma\right)\nonumber\\&+&\frac{1}{2}\Lambda_{v}g_{\omega N}^{2}g_{\rho N}^2\omega_{\mu}\omega^{\mu}\rho_{\mu}\rho^{\mu}
-\frac{1}{4}F_{\mu\nu}F^{\mu\nu}\nonumber\\
&-& \sum_{B}e\overline{\Psi} _{B}\gamma_{\mu}\frac{1+\tau_{3B}}{2}A_{\mu}\Psi_{B}\nonumber\\
&+&\sum_{\ell=e,\mu}{\overline{\Psi}_{\ell}}\left(i\gamma^{\mu}\partial_{\mu} -
M_{\ell}\right) \Psi_{\ell}.
\end{eqnarray}
The equation of motion for baryons, mesons, and photons can be derived from the Lagrangian
density defined in Eq.(\ref{eq:lbm}). The equation of motion for baryons can be given as,

\begin{eqnarray}
\label{eq:dirac}
&\bigg[\gamma^\mu\left(i\partial_\mu - g_{\omega B}\omega_\mu-\frac{1}{2}g_{\rho
B}\tau_{B}.\rho_\mu - e \frac{1+\tau_{3B}}{2}A_\mu \right) - \nonumber\\
&  (M_B + g_{\sigma B}\sigma )\bigg]\Psi_B
=\epsilon_B \Psi_B.
\end{eqnarray}
The Euler-Lagrange equations for the ground-state expectation values of the mesons fields are 

\begin{eqnarray}
\label{eq:sigma}
\left(-\Delta + m_{\sigma}^{2}\right)\sigma & = &\sum_{B} g_{\sigma B}\rho_{sB} -\frac{ \overline{\kappa}}{2} g_{\sigma N}^{3}\sigma^{2}- \frac{\overline{\lambda}}{6} g_{\sigma N}^{4}\sigma ^{3} \nonumber \\ &&+ 
{a_{1}} g_{\sigma N} g_{\omega N}^{2}\omega ^{2}
 +a_{2} g_{\sigma N}^{2}
 g_{\omega N}^{2}\sigma\omega ^{2}
 \nonumber\\&&+ b_{1} g_{\sigma N} g_{\rho B}^{2}\rho ^{2}
 + b_{2} g_{\sigma N}^{2} 
g_{\rho N}^{2}\sigma\rho ^{2}, 
\end{eqnarray}
\begin{eqnarray}
\label{eq:omega}
\left(-\Delta + m_{\omega}^{2}\right)\omega & = &\sum_{B} g_{\omega B}\rho_{B} 
- \frac{\zeta}{6} g_{\omega N}^{4}\omega ^{3} 
\nonumber\\&&-2 a_{1}g_{\sigma N} g_{\omega N}^{2}\sigma\omega 
-a_{2} g_{\sigma N}^{2}
 g_{\omega N}^{2}\sigma^{2}\omega \nonumber\\
&& - \Lambda_{v} g_{\omega N}^{2}
g_{\rho N}^{2}\omega\rho ^{2}, 
\end{eqnarray}
\begin{eqnarray}
\label{eq:rho}
\left(-\Delta + m_{\rho}^{2}\right)\rho & = &\sum_{B} g_{\rho B}\tau_{3B}\rho_{B}- \frac{\xi}{6} g_{\rho N}^{4}\rho ^{3} 
\nonumber\\ &&
-2 b_{1}
 g_{\sigma N} g_{\rho N}^{2}\sigma\rho 
- b_{2} g_{\sigma N}^{2}
 g_{\rho N}^{2}\sigma^{2}\rho \nonumber\\
&& - \Lambda_{v} g_{\omega N}^{2}
g_{\rho N}^{2}\omega^{2}\rho,  
\end{eqnarray}
\begin{equation}
\label{eq:photon}
-\Delta A_{0} = e\rho_{p}.
\end{equation}
where the baryon vector density $\rho_B$,  scalar density $\rho_{sB}$ and charge density
$\rho_{p}$ are, respectively, 
\begin{equation}
\rho_{B}= \left< \overline{\Psi}_B \gamma^0 \Psi_B\right> = \frac{\gamma k_{B}^{3}}{6\pi^{2}},
\end{equation}

\begin{equation}
\rho_{sB} = \left< \overline{\Psi}_B\Psi_B \right> 
          = \frac{\gamma}{(2\pi)^3}\int_{0}^{k_{B}}d^{3}k \frac{M_{B}^*}
            {\sqrt{k^2 + M_{B}^{*2}}},
\end{equation}
\begin{equation}
\rho_{p} = \left< \overline{\Psi}_B\gamma^{0}\frac{1+\tau_{3B}}{2}\Psi_B \right>, 
\end{equation}
 with $\gamma$  the spin-isospin degeneracy. The $M_{B}^{*} = M_{B} - g_{\sigma B}\sigma $
is the effective mass of the baryon species B,  $k_{B}$ is its Fermi momentum
and $\tau_{3B}$ denotes the isospin projections of baryon B.
The energy density of the uniform matter  within the framework of the RMF model is given by;
\begin{equation}
\label{eq:eden}
\begin{split}
{\cal E} & = \sum_{j=B,\ell}\frac{1}{\pi^{2}}\int_{0}^{k_j}k^2\sqrt{k^2+M_{j}^{*2}} dk\\
&+\sum_{B}g_{\omega B}\omega\rho_{B}
+\sum_{B}g_{\rho B}\tau_{3B}\rho_{B}\rho
+ \frac{1}{2}m_{\sigma}^2\sigma^2\\
&+\frac{\overline{\kappa}}{6}g_{\sigma N}^3\sigma^3
+\frac{\overline{\lambda}}{24}g_{\sigma N}^4\sigma^4
-\frac{\zeta}{24}g_{\omega N}^4\omega^4\\
&-\frac{\xi}{24}g_{\rho N}^4\rho^4
 - \frac{1}{2} m_{\omega}^2 \omega ^2
-\frac{1}{2} m_{\rho}^2 \rho ^2\\
&-a_{1} g_{\sigma N}
 g_{\omega N}^{2}\sigma \omega^2
 -\frac{1}{2}a_{2} g_{\sigma N}^2 g_{\omega N}^2\sigma^2 \omega^2\\
 &-b_{1}g_{\sigma N}g_{\rho N}^2 \sigma\rho^2
-\frac{1}{2} b_{2} g_{\sigma N}^2 g_{\rho N}^2\sigma^2\rho^2\\
 &- \frac{1}{2} \Lambda_{v} g_{\omega N}^2 g_{\rho N}^2
\omega^2\rho^2.\\
\end{split}
\end{equation}
The pressure of the uniform matter  is given by
\begin{equation}
\label{eq:pden}
\begin{split}
P & = \sum_{j=B,\ell}\frac{1}{3\pi^{2}}\int_{0}^{k_j}
\frac{k^{4}dk}{\sqrt{k^2+M_{j}^{*2}}} 
- \frac{1}{2}m_{\sigma}^2\sigma^2\\
&-\frac{\overline{\kappa}}{6}g_{\sigma N}^3\sigma^3 
-\frac{\overline{\lambda}}{24}g_{\sigma N}^4\sigma^4
+\frac{\zeta}{24}g_{\omega N}^4\omega^4\\
&+\frac{\xi}{24}g_{\rho N}^4\rho^4
  + \frac{1}{2} m_{\omega}^2 \omega ^2
+\frac{1}{2} m_{\rho}^2 \rho ^2 \\
& +a_{1} g_{\sigma N}
g_{\omega N}^{2}\sigma \omega^2
+\frac{1}{2} a_{2} g_{\sigma N}^2 g_{\omega N}^2\sigma^2 \omega^2\\
&+b_{1}g_{\sigma N}g_{\rho N}^2 \sigma\rho^2
+\frac{1}{2} b_{2} g_{\sigma N}^2 g_{\rho N}^2\sigma^2\rho^2\\
&+ \frac{1}{2} \Lambda_{v} g_{\omega N}^2 g_{\rho N}^2
\omega^2\rho^2.\\
\end{split}
\end{equation}
Here, the sum is taken over nucleons and leptons.
\begin{table*}
\centering
\caption{\label{tab:table1}
Newly generated parameter sets HPU1, HPU2 and HPU3  of RMF  Lagrangian given
in Eq.(\ref{eq:lbm}) alongwith the theoretical uncertainties/errors.
The parameters $\overline{\kappa}$, $a_{1}$, and $b_{1}$ are
 in  fm$^{-1}$.
 The mass for nucleon, $\omega$ and $\rho$ meson is taken as $M_N$ =939 MeV, $m_{\omega}$ = 782.5 MeV and $m_{\rho}$ = 770 MeV .
The values of $\overline{\kappa}$, $\overline{\lambda}$, $a_{1}$, $a_{2}$, $b_{1}$ and
${b_{2}}$
 are multiplied by $10^{2}$. Parameters for DOPS1, NL3 and Big Apple models are also shown for comparison.}
\vskip 1cm
\begin{tabular}{ccccccc}
\hline
\hline
\multicolumn{1}{c}{${\bf {Models}}$}&
\multicolumn{1}{c}{{\bf HPU1}}&
\multicolumn{1}{c}{{\bf HPU2}}&
\multicolumn{1}{c}{{\bf HPU3}}&
\multicolumn{1}{c}{{\bf DOPS1}}&
\multicolumn{1}{c}{{\bf NL3}}&
\multicolumn{1}{c}{{\bf Big Apple}}\\
\hline
${\bf g_{\sigma}}$  &9.37959$\pm$0.03534&9.91463$\pm$0.06587 &9.87733$\pm$0.06297&10.20651&10.21743&9.67810\\
${\bf g_{\omega}}$  &11.63792$\pm$0.06056& 12.45333$\pm$0.04816&12.45229$\pm$0.05479&12.87969&12.86762&12.33541\\
${\bf g_{\rho}}$ &10.79751$\pm$0.80596&10.65758$\pm$1.07023 &11.34072$\pm$2.48071&14.13399&8.94800&14.14256\\
${\bf \overline {\kappa}}$&3.14479$\pm$0.17668&2.45494$\pm$0.12517 &3.10148$\pm$0.53829 &2.62033&1.95734&2.61776\\
${\bf \overline {\lambda}}$&-2.51218$\pm$0.25041&-1.67119$\pm$0.18478&-1.57402$\pm$0.33738& -1.67616&-1.59137&-2.16586\\
${\bf{\zeta}}$&-&0.00682$\pm$0.23594&0.003301$\pm$0.03266&- &-&0.000699\\
${\bf \Lambda_{v}}$ &0.05396$\pm$0.02277&0.04745$\pm$0.02773 &0.00248$\pm$0.47194 &0.00869&-&0.09400\\
${\bf a_1}$&-&- &0.03559$\pm$0.03112 &002169&-&-\\
${\bf  a_2}$  &- &- &0.01428$\pm$0.16366 &0.01785&-&-\\
${\bf b_1}$ &- &- &0.47507$\pm$1.78695 &0.73554&-&-\\
${\bf b_2}$ &-&- &0.92769$\pm$0.34847 &0.98545&-&-\\
${\bf m_{\sigma}}$ &497.976$\pm$3.004&501.606$\pm$4.346 &498.638$\pm$5.172 &503.620&508.194&492.975\\
\hline
\hline
\end{tabular}
\end{table*}
\section{parametrization of RMF Model}\label{parameter}
The optimization of the parameters (p) appearing in the Lagrangian (Eq. \ref{eq:lbm}) has been performed  by using the simulated annealing method (SAM) \cite{Burvenich2004,
Kirkpatrick1984} by following $\chi^{2}$ minimization procedure  which is given
as,
\begin{equation}
	{\chi^2}(\textbf{p}) =  \frac{1}{N_d - N_p}\sum_{i=1}^{N_d}
\left (\frac{O{_i^{exp}} - O_i^{th}}{\sigma_i}\right )^2 \label {chi2},
\end{equation}
where $N_d$ is the number of  experimental data points and $N_p$ is the number
of fitted parameters. The $\sigma_i$ denotes adopted errors \cite{Dobaczewski2014}
and $O_i^{exp}$ and $O_i^{th}$ are the experimental and the corresponding
theoretical values, respectively, for a given observable.
We search the parameters of the model by fitting the  available experimental data of total binding energies  and  charge rms radii \cite{Wang2017,Wang2021,Otten1989,Vries1987} for some closed/open-shell nuclei $^{16,24}$O,$^{40,48,54}$Ca, $^{56,68,78}$Ni,$^{88}$Sr,$^{90}$Zr,$^{100,116,132,138}$Sn,  and $^{144}$Sm, $^{208}$Pb.  We have also included the recently measured neutron skin thickness ($\Delta r_{np}$) \cite{Adhikari2021} in our fit data to constrain the linear density dependence of symmetry energy. The maximum mass of neutron star M$_{max}$= 2.35$\pm$0.17 M$_{\odot}$ \cite{Romani2022} is also incuded in the fitting protocol. The pairing has been included for the open shell nuclei by using the  BCS formalism with constant pairing gaps that have been taken from the particle separation energies of neighboring nuclei \cite{Ring1980,Karatzikos2010,Wang2021}.
The parameter sets are generated in consideration of pulsar PSR J0952-0607 also satisfying  finite and bulk nuclear matter properties. The heaviest observed neutron star mass for the black widow pulsar and PREX-II results demands a stiff EoS and a stiff slope of the symmetry energy coefficient.  \\
We generate three parameter sets for different combinations of  non-linear, self and cross-couplings among isoscalar-scalar $\sigma$, isoscalar-vector $\omega_{\mu}$ and isovector-vector $\rho_{\mu}$ meson fields up to the quartic order as enumerated below.\\
1. In the HPU1 parametrization, self interactions ($\overline {\kappa}, \overline{\lambda}$) of $\sigma$ meson and cross interaction term ($\Lambda_{v}$) of $\omega^{2}$-$\rho^{2}$ mesons, in addition to the exchange interactions of baryons with $\sigma$, $\omega$ and $\rho$ mesons are taken in the Lagrangian. The $\omega$ meson self-interaction  term  $\zeta$ is not included in order to maintain compatibility with  heaviest observed  neutron star mass  M$_{max}$= 2.35 $\pm$0.17 M$_{\odot}$ for black widow pulsar PSR J0952-0607.\\
2. In the  HPU2 parametrization,  we  also incorporate the $\omega$ meson self coupling parameter $\zeta$ in addition to the coupling terms considered in HPU1 model.\\
3. In the HPU3 parametrization, we include all possible self and cross-couplings among isoscalar-scalar $\sigma$, isoscalar-vector $\omega_{\mu}$ and isovector-vector $\rho_{\mu}$ meson fields up to the quartic order so that this parameter set  satisfy the  mass constraints of PSR J0952-0607 along with finite nuclear properties and  PREX-II results on neutron skin thickness of $^{208}$Pb. The inclusion of these possible self and cross interactions terms of $\sigma$, $\omega$ and $\rho$ mesons are important to accommodate naturalness behavior of parameters as imposed by effective field theory \cite{Furnstahl1997}. In Table \ref{tab:table1}, we display the values of coupling parameters for the HPU1, HPU2 and HPU3 parametrizations generated for the Lagrangian given by Eq. (\ref{eq:lbm}) along with the theoretical uncertainties/errors  calculated by the method discussed in Refs. \cite{Brandt1997, Dobaczewski2014}. The values of parameter sets  for DOPS1 \cite{Virender2022a}, NL3 \cite{Lalazissis1997}  and Big Apple \cite{Fattoyev2020}  are also shown.\\
\begin{figure*}
\includegraphics[trim=0 0 0 0,clip,scale=0.5]{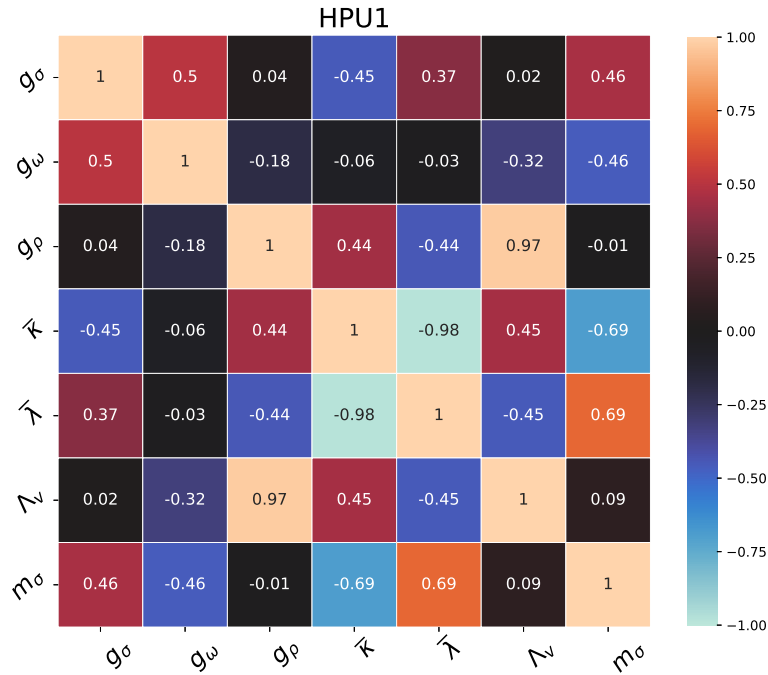}
\caption{\label{par_hpu1} (color online)  Correlation coefficients   among the model parameters
of the Lagrangian given by Eq. (\ref{eq:lbm}) for HPU1 parametrization.  }
\end{figure*}
\begin{figure*}
\includegraphics[trim=0 0 0 0,clip,scale=0.5]{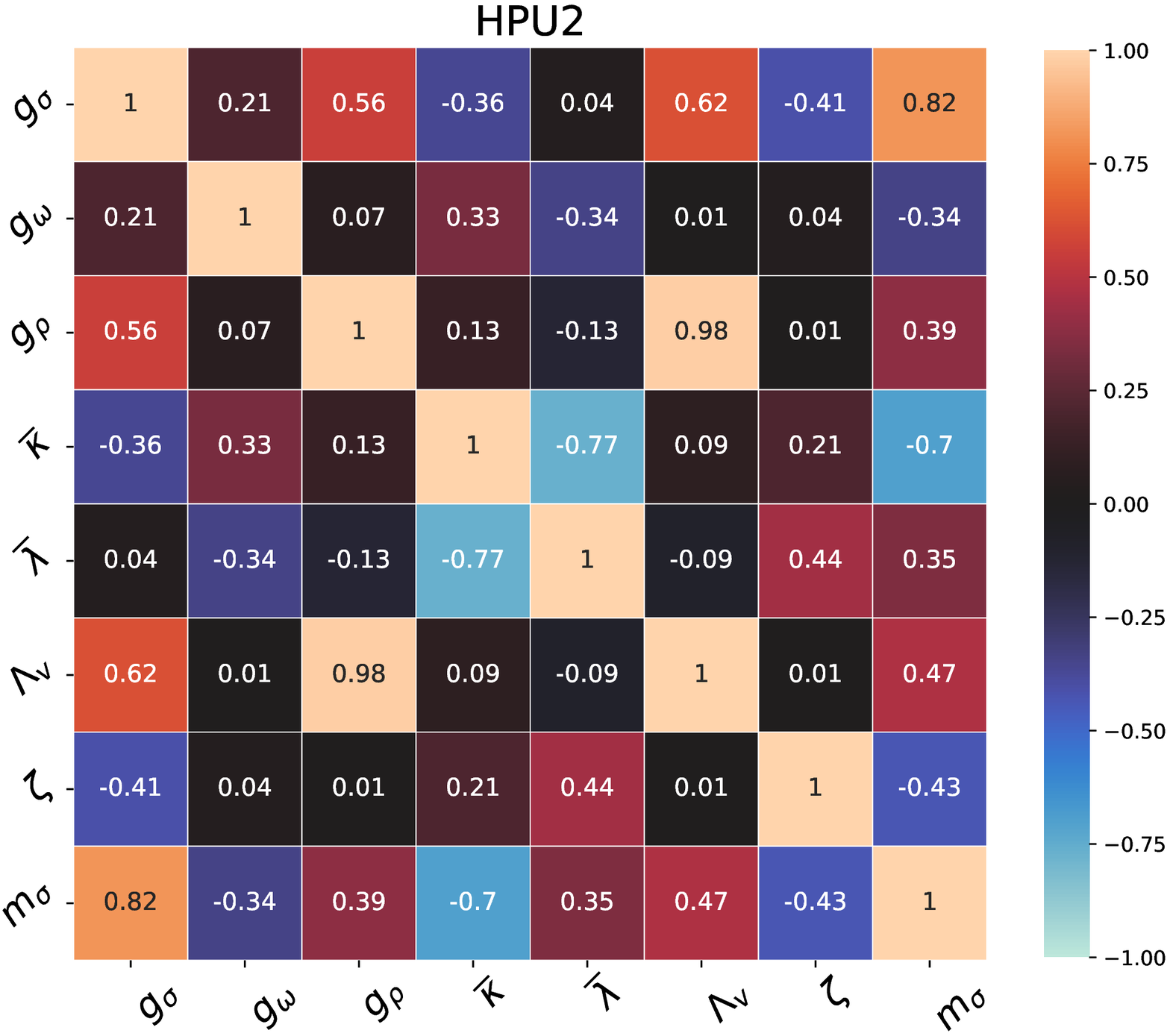}
\caption{\label{par_hpu2} (color online) Same as Fig.\ref{par_hpu1}, but for HPU2 parametrization.}
\end{figure*}
\begin{figure*}
\includegraphics[trim=0 0 0 0,clip,scale=0.5]{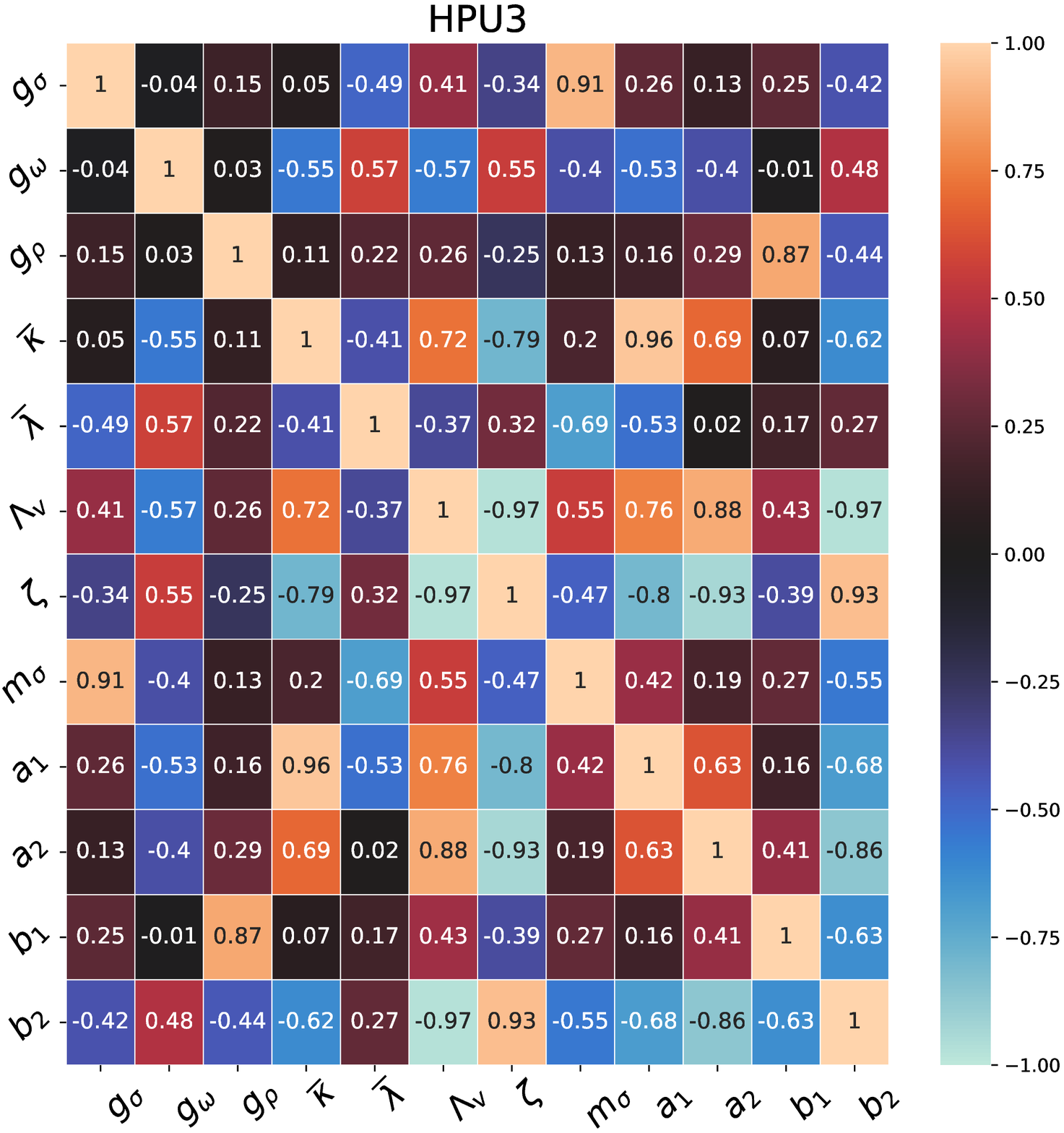}
\caption{\label{par_hpu3} (color online) Same as Fig.\ref{par_hpu1}, but for for HPU3 parametrization). }
\end{figure*}
Having obtained the parameters sets, we also calculate the correlations coefficients between two parameters/observables by the covariance analysis as discussed in Refs. \cite{Brandt1997, Fattoyev2011, Dobaczewski2014,Mondal2015}.
In Figs. \ref{par_hpu1}, \ref{par_hpu2} and \ref{par_hpu3} we show the plots for the correlation coefficients between  the coupling parameters appearing in Lagrangian (Eq.\ref{eq:lbm}) for HPU1, HPU2 and HPU3 models.
It can be observed from Figs. \ref{par_hpu1} and \ref{par_hpu2} that a strong correlation exists between the pairs of coupling parameters $\Lambda_{v}$ - $g_{\rho}$, and $\overline{\kappa}-\overline{\lambda}$ for HPU1 and HPU2 models. In addition, a strong  correlation is also observed for pair of coupling parameters $g_{\sigma}$ - m$_{\sigma}$ for HPU2 model. It can be observed that for HPU3 model, coupling parameter $\Lambda_{v}$ shows a strong dependence on cross interaction terms. The correlation coefficient  between $\Lambda_{v}$ and $g_\rho$ becomes weak and $\Lambda_{v}$ shows a strong correlations with coupling parameters a$_{1}$, a$_{2}$ and   b$_{2}$ and a good correlations with b$_{1}$. The isovector coupling parameter $g_{\rho}$ is found to be strongly correlated with b$_{1}$. A strong correlation between the model parameters indicates a strong interdependence i.e. if one parameter is fixed at a certain value then the other must attain the precise value as suggested by their correlation. The effective field theory imposes the condition of naturalness  \cite{Furnstahl1997} on the parameters or expansion coefficients appearing in the effective Lagrangian density  Eq. (\ref{eq:lbm}). According to naturalness, the coefficients of various terms in Lagrangian density functional should be of the same size when expressed in an appropriate dimensionless ratio. The dimensionless ratios are obtained by dividing Eq. (\ref{eq:lbm}) by $M^{4}$ and expressing each term in powers of $\frac{g_{\sigma}\sigma}{M}$, $\frac{g_{\omega}\omega}{M}$ and 2$\frac{g_{\rho}\rho}{M}$. This indicates that  the dimensionless ratios ${\frac{1}{2C_{\sigma}^{2}M^{2}}}$,${\frac{1}{2C_{\omega}^{2}M^{2}}}$, ${\frac{1}{8C_{\rho}^{2}M^{2}}}$, ${\frac{\overline{\kappa}}{6M}}$, ${\frac{\overline{\lambda}}{24}}$, ${\frac{{\zeta}}{24}}$, ${\frac{a_1}{M}}$, ${\frac{a_2}{2}}$, ${\frac{b_1}{4M}}$, ${\frac{b_2}{8}}$ and ${\frac{\Lambda_{v}}{8}}$ should be roughly of same size, where ${C_{i}}^{2}=\frac{{g_{i}}^{2}}{{M_{i}}^{2}}$, i denotes $\sigma$, $\omega$ and $\rho$ mesons.
\begin{table*}
\centering
\caption{\label{tab:nat}
The values of parameters are expressed as dimensionless ratios corresponding to naturalness behavior. All  values have been  multiplied by $10^{3}$.}
\begin{tabular}{cccccccc}
\hline
\hline
\multicolumn{1}{c}{{ {Parameters}}}&
\multicolumn{1}{c}{{ HPU1}}&
\multicolumn{1}{c}{{ HPU2}}&
\multicolumn{1}{c}{{ HPU3}}&
\multicolumn{1}{c}{{DOPS1}}&
\multicolumn{1}{c}{{NL3}}&
\multicolumn{1}{c}{{ Big Apple}}\\
\hline
${\bf{\frac{1}{2C_{\sigma}^{2}M^{2}}}}$ & 1.598&1.451 &1.445&1.381&1.403&1.469\\
${\bf{\frac{1}{2C_{\omega}^{2}M^{2}}}}$  &2.564& 2.238&2.239&2.093&2.097&2.282\\
${\bf{\frac{1}{8C_{\rho}^{2}M^{2}}}}$ &0.721 &0.7406&0.654&0.421 &1.031&0.412\\
${\bf{\frac{\overline{\kappa}}{6M}}}$&1.101 &0.860 &1.086 &0.918&0.668&0.917\\
${\bf{\frac{\overline{\lambda}}{24}}}$&-1.047&-0.696&-0.656& -0.698&-0.663&-0.902\\
${\bf{\frac{{\zeta}}{24}}}$&-&0.284&0.138 & -&-&0.029\\
${\bf{\frac{a_1}{M}}}$ &- &- &0.356 &0.217&-&-\\
${\bf{\frac{a_2}{2}}}$&- &- &0.101 &0.089&-&-\\
${\bf{\frac{b_1}{4M}}}$&- &- &1.187 &1.839&-&-\\
${\bf{\frac{b_2}{8}}}$ &- &- &1.159 &1.232&-&-\\
${\bf{\frac{\Lambda_{v}}{8}}}$ &6.745 &5.931 &0.310 &1.086&-&11.750\\
\hline
\hline
\end{tabular}
\end{table*}
In Table \ref{tab:nat}, we display the overall naturalness behavior of HPU's parameterizations  i.e. the value of these coupling parameters when expressed in appropriate dimensionless ratios as discussed above. The corresponding values for DOPS1, NL3,  and Big Apple parameter sets are also shown for sake of comparison. It is obvious from the table that HPU3  parameterization closely favors the naturalness behavior.  This may be attributed to the fact that this parameterization includes  possible self and crossed interaction terms of $\sigma$, $\omega$, and $\rho$-mesons up to the quartic order.
 The small value of parameter $\Lambda_{v}$ (cross interaction term of $\omega^{2}$ and $\rho^{2}$) appearing in Eq. (\ref{eq:lbm})  for HPU3 model which might be responsible for  better naturalness behaviour of the parameters is  attributed to the fact that the coupling parameter $\Lambda_{v}$ shows strong dependence on  cross coupling terms $a_{1}$, $a_{2}$,  $b_{1}$and $b_{2}$ as suggested by their correlation coefficents (see Fig. \ref{par_hpu3}).
It is evident from the Table \ref{tab:table1} that the value of coupling parameter $\Lambda_{v}$ is relatively larger  for HPU1, HPU2 and Big Apple and shows deviation from the naturalness behavior  might be attributed to the fact of not including cross interactions in their respective Lagrangian. It can be seen from Table \ref{tab:nat} that the value of coupling term $\Lambda_{v}$ after expressed in appropriate  dimensionless ratio   is 6.745, 5.931 and 11.750 for HPU1, HPU2 and Big Apple models respectively.
  Keeping in view the naturalness behavior of the parameters as imposed by the effective field theory \cite{Furnstahl1997} and as observed  in HPU3 model, it can be concluded that cross intercations terms of mesons have significant role and their contributions have to be incorporated in the Lagrangian. The naturalness behavior of parameters can be further improved by considering the next higher order terms containing the gradient of fields   \cite{Furnstahl1997}. NL3 parameterization favors  the naturalness behavior  of the parameter but it does not include any cross interaction terms of  $\omega$, and $\rho$ mesons which are very important for constraining the symmetry energy and its density dependence. Also, it is observed from the table that the $\omega$ meson coupling term $\zeta$ should be either zero or attain a  a very small value in the calibration procedure of parameters in order to maintain compatibility with the high maximum mass of neutron stars like pulsar PSR J0952-0607.
\section{RESULTS AND DISCUSSION}\label{results}
In this section we use the model parametrizations HPUs to calculate the properties of finite nuclei, bulk nuclear matter and neutron stars. We also discuss the correlations among nuclear matter observables and model
parameters.
\subsection{Properties of finite nuclei and nuclear matter }
The newly generated  parametrizations HPU1, HPU2 and HPU3  give a good fit to the properties of finite nuclei. The binding energies obtained using HPUs  parametrizations are in harmony with the available experimental data   \cite{Wang2017,Wang2021}. The value of rms errors in the total binding energies  calculated for HPU1, HPU2 and HPU3  parametrizations are  3.06, 1.81 and 2.35 MeV respectively. The root mean square (rms) errors in  charge radii for all nuclei taken in our fit are 0.050, 0.016 and 0.017 fm for HPUs parameter sets respectively. The neutron skin thickness  for $^{208}$Pb   comes out to be 0.216, 0.218 and 0.217 fm for HPU1, HPU2 and HPU3 parametrizations respectively and is in close proximity with the limits put by PREX-II results \cite{Adhikari2021}. The value of neutron skin thickness  of $\Delta r_{np} (^{48}{\rm Ca})$ obtained for these parametrizations is consistent with the results reported in Ref.\cite{Reinhard2022a}.\\
In Table \ref{tab:table2}, we present the results for the  SNM properties such as binding energy per nucleon (E/A), incompressibility (K), the  effective nucleon mass ($M^*$)   at the saturation density ($\rho_{0}$),   symmetry energy coefficient (J), the slope of  symmetry energy (L) and curvature parameter  $K_{\rm sym}$. These SNM properties play an important role for constructing the EOS for nuclear matter. The results  are also compared with the DOPS1 \cite{Virender2022a}, NL3 \cite{Lalazissis1997} and Big Apple \cite{Fattoyev2020} parameter sets.
\begin{table*}
\caption{\label{tab:table2}
The bulk nuclear matter properties (NMPs) at saturation density  for HPUs parametrization compared with that other parameter sets. $\rho_{0}$,  E/A,  K, $M^{*}/M$, J,  L and $K_{sym}$  denote the saturation density, Binding Energy per nucleon, Nuclear Matter incompressibility coefficient, the ratio of effective nucleon mass to the nucleon mass, Symmetry Energy, the slope of symmetry energy, and curvature of symmetry energy   respectively. the value of $\rho_0$ is in fm$^{-3}$ and rest all the quantities are in MeV. The values of neutron skin thickness $\Delta r_{np}$ for $^{208}$Pb and $^{48}$Ca nuclei in units of fm are also listed.}
\vskip 1cm
\begin{tabular}{ccccccc}
\hline
\hline
\multicolumn{1}{c}{${ {NMPs}}$}&
\multicolumn{1}{c}{{ HPU1}}&
\multicolumn{1}{c}{{ HPU2}}&
\multicolumn{1}{c}{{ HPU3}}&
\multicolumn{1}{c}{{ DOPS1}}&
\multicolumn{1}{c}{{ NL3}}&
\multicolumn{1}{c}{{ Big Apple}}\\
\hline
${\bf{\rho_{0} }}$&0.157&0.150 &0.151&0.150&0.148&0.155\\
${\bf E/A}$ &-16.298 & -16.119&-16.062&-16.073&-16.248&-16.339\\
${\bf K}$ &229.88&225.84 &225.56&231.20&271.56&227.09\\
${\bf ~M^{*}/M}$&0.639&0.621&0.635& 0.604&0.595&0.608\\
${\bf J}$&34.34&33.21&33.22&31.89&37.40&31.41\\
${\bf L}$&61.21&63.87&75.03& 65.59&118.56&40.34\\
{${\bf {K_{sym}} }$}&-96.85&-70.78&-31.42&18.01&100.90&89.58\\
{${\bf {\Delta r_{np}~ (^{208}Pb)} }$}&0.216&0.218&0.217&0.185&0.279&0.150\\
${\bf {\Delta r_{np} ~(^{48}Ca)}}$&0.202&0.201&0.203&0.182&0.226&0.168\\
\hline
\hline
\end{tabular}
\end{table*}
The E/A lies in the range  of 16.062 - 16.298 MeV for HPUs parametrization. The  value of J and L obtained by HPUs parametrizations are  consistent with the values J = 38.1 $\pm$ 4.7 MeV and L = 106 $\pm$ 37 MeV as  inferred by Reed et al., \cite{Reed2021} and is also consistent with the constraints from the observational analysis J = 31.61 $\pm$ 2.66 and L = 58.9 $\pm$ 16 MeV \cite{Li2013}. The slope of symmetry energy obtained for HPU1 parameter set also satisfies the recently reported limit  L = 54 $\pm$ 8 MeV \cite{Reinhard2021} and
L= 15.3$^{+46.8}_{-41.5}$
\cite{Zhang2022}. The value of K is in the range 225.56 -229.88  MeV  which is in agreement with the value K = 240 $\pm$ 20 MeV determined from isoscalar giant monopole resonance (ISGMR) for  $^{90}$Zr and $^{208}$Pb nuclei \cite{Colo2014,Piekarewicz2014}. The value of curvature of symmetry energy K$_{sym}$ for HPUs parameter sets also satisfies the empirical limit discussed in \cite{Zimmerman2020}. The value of $K_{sym}$ is determined only poorly \cite{Newton2021,Xu2022,Gil2022} and the experimental data on finite nuclei is not enough to  constrain $K_{\rm sym}$. Only the accurate knowledge of symmetry energy at higher densities ($\rho > 2\rho_0$)   may constrain the $K_{\rm sym}$ in tighter bounds. This may be attributed to the large experimental error on the neutron skin thickness for  $^{208}$Pb (0.283 $\pm$ 0.071 fm) which lead us to choose the large adopted error during the optimization procedure. The values of neutron-skin thickness $(\Delta r_{np})$ for $^{208}$Pb and $^{48}$Ca nuclei are also displayed
in Table \ref{tab:table2}.  It can be observed from the Tables \ref{tab:table1} and \ref{tab:table2} that the slope of symmetry energy (L)  has strong dependence on cross interactions term ($\Lambda_{v}$) of $\omega^{2}$ - $\rho^{2}$ mesons. The value of L decreases with increase in $\Lambda_{v}$. As the value of coupling $\Lambda_{v}$ increases from 0.00248 (HPU3) to 0.09400 (Big Apple), the corresponding value of L decreases from 75.03 MeV to 40.34 MeV. For NL3 parameter set, the value of L is large (118.56 MeV), may be due to not including the cross interaction term $\Lambda_{v}$. The value of symmetry energy coefficient J at 2$\rho_{0}$ for HPU1, HPU2, HPU3 are found to be 53.15, 53.40 and 57.59 MeV respectively and is consistent with the constraints  J(2$\rho_{0}$)= 51$\pm$13 MeV inferred from  nine new analyses of neutron star observables since GW170817 \cite{Li2021} and J(2$\rho_{0}$ )= 62.8$\pm$15.9 MeV \cite{Yue2022}. The value of J at 2$\rho_{0}$ for HPU1 and HPU2 models are very close to the constraint J(2$\rho_{0}$) = 40.2$\pm12.8$ MeV based on microscopic calculations with various energy density functionals \cite{Chen2015}. The value J at 2$\rho_{0}$ for DOPS1, NL3 and Big Apple models comes out to be  55.23, 78.31 and 49.96 MeV respectively. Like the slope of symmetry energy (L),  J  at 2$\rho_{0}$ also  shows similar trend with cross interaction couplng  $\Lambda_{v}$.

\begin{figure}
\includegraphics[trim=0 0 0 0,clip,scale=0.48]{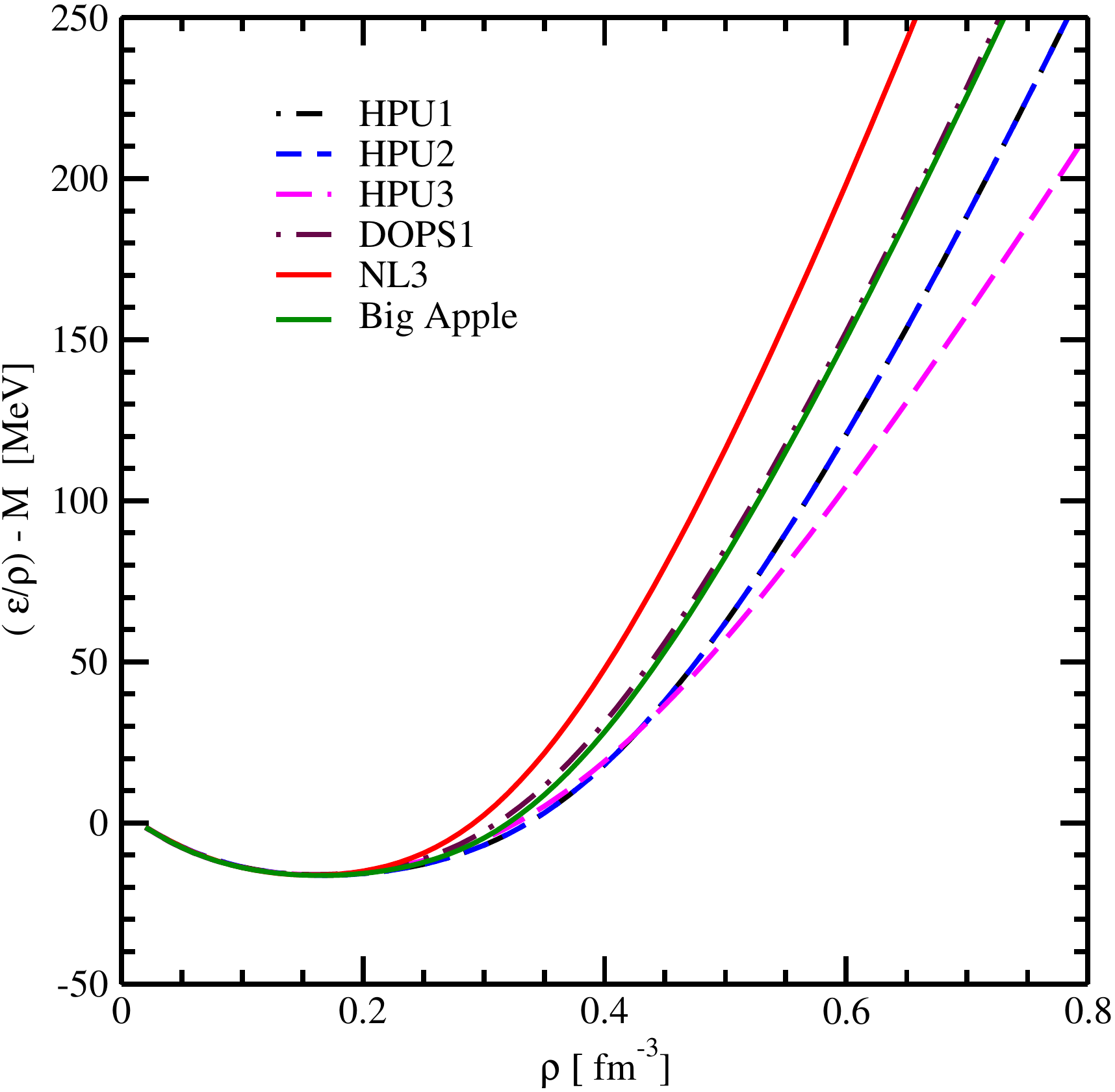}
\caption{\label{ea} (color online) Binding energy per nucleon in the symmetric nuclear matter as a function of baryon density Variation of Pressure as a function of baryon density for various parametrizations.}
\end{figure}
\begin{figure*}
\includegraphics[trim=0 0 0 0,clip,scale=0.48]{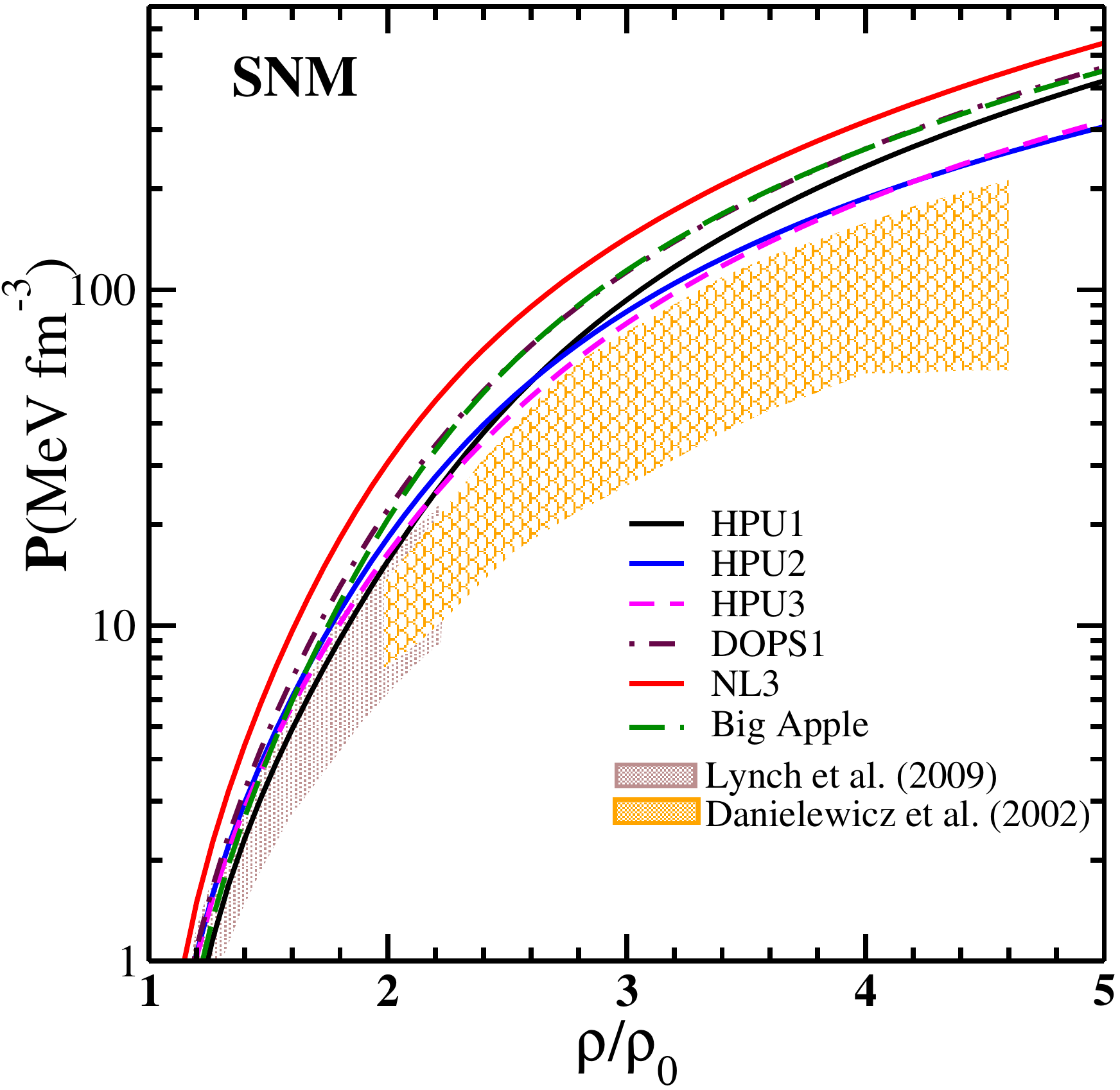}
\caption{\label{snm} (color online) Pressure as a function of baryon density for symmetric nuclear matter calculated with  HPUs  parametrizations along with DOPS1,NL3 and Big Apple models. The shaded regions represent the experimental data taken from the references \cite{Danielewicz2002,Lynch2009}.}
\end{figure*}
\begin{figure*}
\includegraphics[trim=0 0 0 0,clip,scale=0.48]{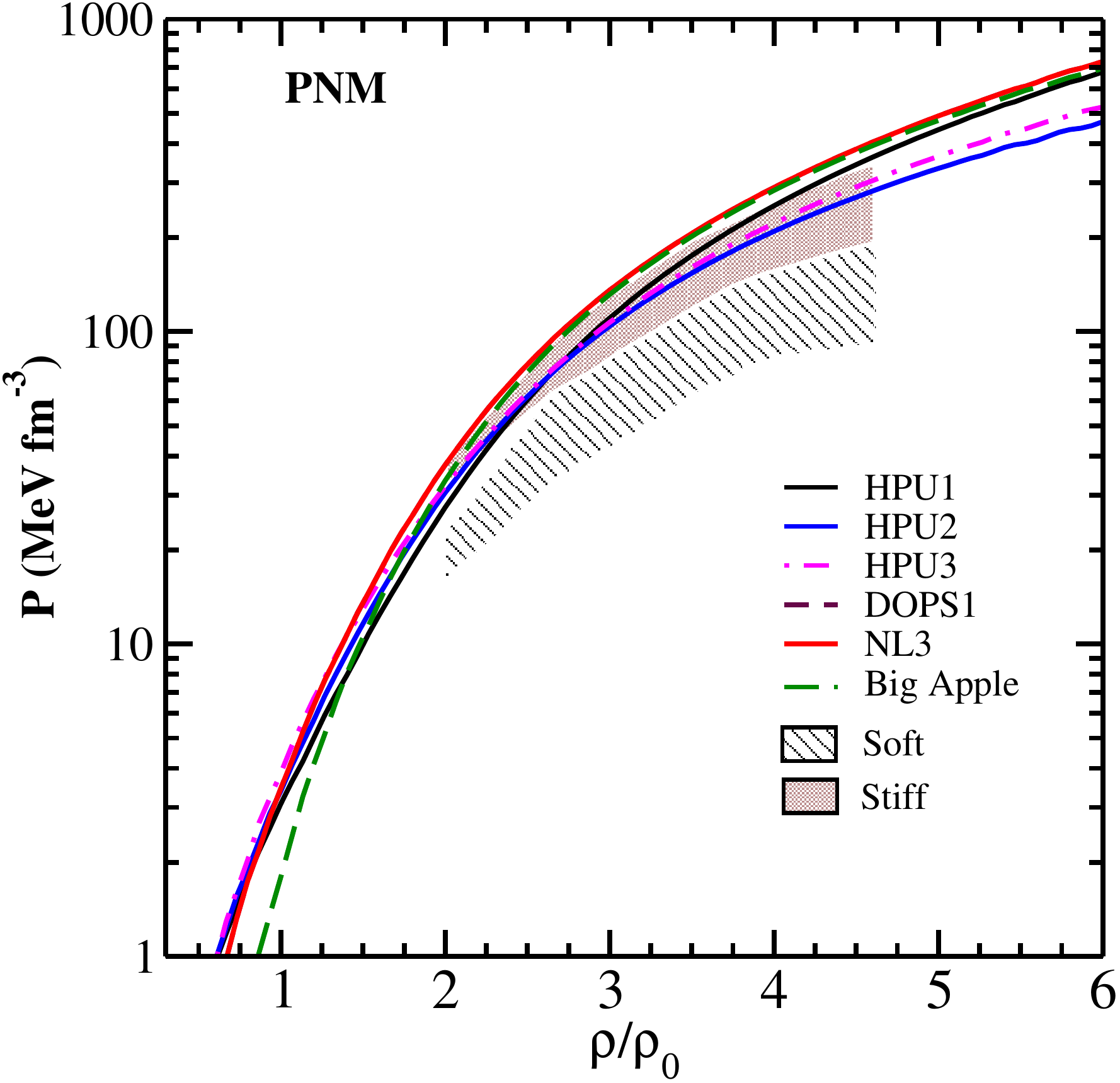}
\caption{\label{pnm} (color online) Variation of Pressure as a function of baryon density for pure neutron matter computed with  HPUs parametrizations along with DOPS1, NL3 and Big Apple models. The shaded region represents the experimental data taken from the reference \cite{Danielewicz2002}.}
\end{figure*}
In Fig.\ref{ea}, we illustrate the binding energy per nucleon  E$_{B}$ = ($ {\cal E}/{\rho}$ )- M,  as a function of baryon density for symmetric nuclear matter for various parametrizations. E$_{B}$ near saturation density is almost similar for all models considered here but it shows significant difference  at higher densities. As it can be seen from the figure that HPU1, DOPS1 and NL3  parametrizations show stiff behavior (large E$_{B}$) whereas HPU2 and HPU3 parameters show soft behavior (small E$_{B})$ at higher density regions. It may be due to the absence and presence of $\omega$ meson coupling parameter in the Lagrangian of the respective parametrization.
In Figs.(\ref{snm} and \ref{pnm}), we plot the EoS i.e. pressure as a function of baryon density  scaled to  the saturation density  for SNM  and  PNM  using the HPUs parametrizations. Similar results are also shown for NL3, DOPS1 and Big Apple models. The shaded regions represent the experimental data taken from the Refs. \cite{Danielewicz2002,Lynch2009}. The EOS calculated by using the HPUs parametrizations are relatively stiffer which is in requirement to constrain the recent astrophysical observations \cite{Romani2022}. It is evident from Fig. \ref{snm} that in the low density regime (up to 2$\rho_{0}$), the EoSs for SNM obtained for HPUs parametrizations are close to the upper limit of the allowed region with the EoS extracted from the analysis of the particle flow in heavy ion collision \cite{Danielewicz2002} and lie in the upper portion of  the allowed region of EoS extracted from Ref. \cite{Lynch2009}. After 2$\rho_{0}$, the SNM pressure for all HPU's  parameterizations considered in the present work start deviating from the collective flow constraining band. This might be attributed to the fact that  these parmetrizations are  obtained keeping in view the heaviest observed neutron star M$_{max}$ = 2.35$\pm$ 0.17 M$_{\odot}$ for the black widow pulsar PSR J0952-0607 and  demands stiff EoSs i.e. larger value of pressure at higher densities.  Also, it can be seen from Fig.\ref{pnm} that EoSs obtained for PNM  using HPUs parameter sets  lie in the allowed stiff region. The HPUs parametrizations demand stiff EoSs, whereas the prediction of neutron star maximum mass around $2M_{\odot}$ and constraints on EOSs of SNM and PNM  as extracted from the analysis of particle flow in heavy ion collisions \cite{Danielewicz2002} require relatively softer EOSs as demanded by  GW170817 event.  \cite{Danielewicz2002}. It is evident from the figures that the EOSs for SNM and PNM calculated with the NL3, DOPS1 and Big Apple  parametrizations are very stiff and follow the similar behavior as follwed by HPUs models.\\
\begin{figure*}
\includegraphics[trim=0 0 0 0,clip,scale=0.68]{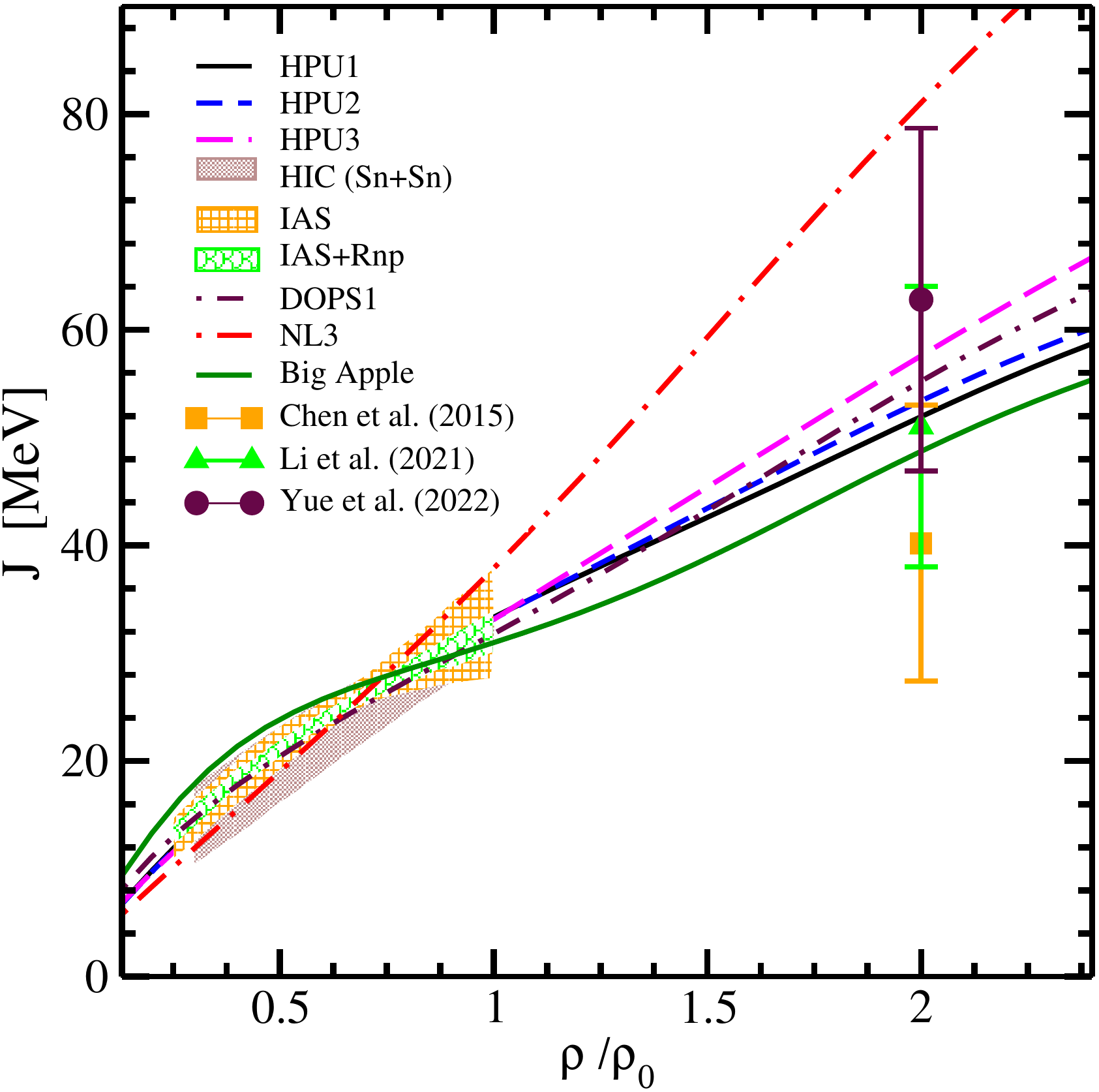}
\caption{\label{jrho} (color online) Symmetry energy  as a function of baryon density for various models considered in the present work. The shaded regions represent the constraints on density dependence of symmetry energy from heavy ion collisions and isobaric analysis states (IAS) taken from Refs.  \cite{Tsang2009,Danielewicz2014}. The constraints on  magnitude of symmetry energy coefficient  at J(2$\rho_{0}$): J(2$\rho_{0}$)= 62.8$\pm$15.9 MeV \cite{Yue2022}, J(2$\rho_{0}$)= 51$\pm$13 MeV from nine new analyses of neutron star observables since GW170817 \cite{Li2021} and  J(2$\rho_{0}$) = 40.2$\pm12.8$ MeV based on microscopic calculations with various energy density functionals \cite{Chen2015} are also shown. }
\end{figure*}
In Fig. \ref{jrho}, we plot the symmetry energy as a function of baryon density for various HPUs  parametrizations. The results for  other models   are also shown for comparison. The shaded regions represent the constraints on density dependence of symmetry energy from heavy ion collisions and isobaric analogue states (IAS) taken from Refs. \cite{Tsang2009,Danielewicz2014}. The constraints on  magnitude of symmetry energy coefficient  at J(2$\rho_{0}$): J(2$\rho_{0}$)= 62.8$\pm$15.9 MeV \cite{Yue2022}, J(2$\rho_{0}$)= 51$\pm$13 MeV from nine new analyses of neutron star observables since GW170817 \cite{Li2021} and  J(2$\rho_{0}$) = 40.2$\pm12.8$ MeV based on microscopic calculations with various energy density functionals \cite{Chen2015} are also shown. It can be observed that the symmetry energy increases with  baryon density  for various models considered in the present work and also lie in the allowed regions and satisfies various constraints as discussed above. Amongst HPUs parametrizations, the value of J is found to be stiffest for HPU3 and softest for HPU1 model in higher density regime. This might be due to the smaller value of coupling $\Lambda_{v}$ in case of HPU3 as compared to its value for HPU1 as this coupling term plays an important role for constraining the symmetry energy and its density dependence.
\subsection{Neutron star properties}
\begin{figure*}
\hspace{-0.3cm}\includegraphics[trim=0 0 0 0,clip,scale=0.5]{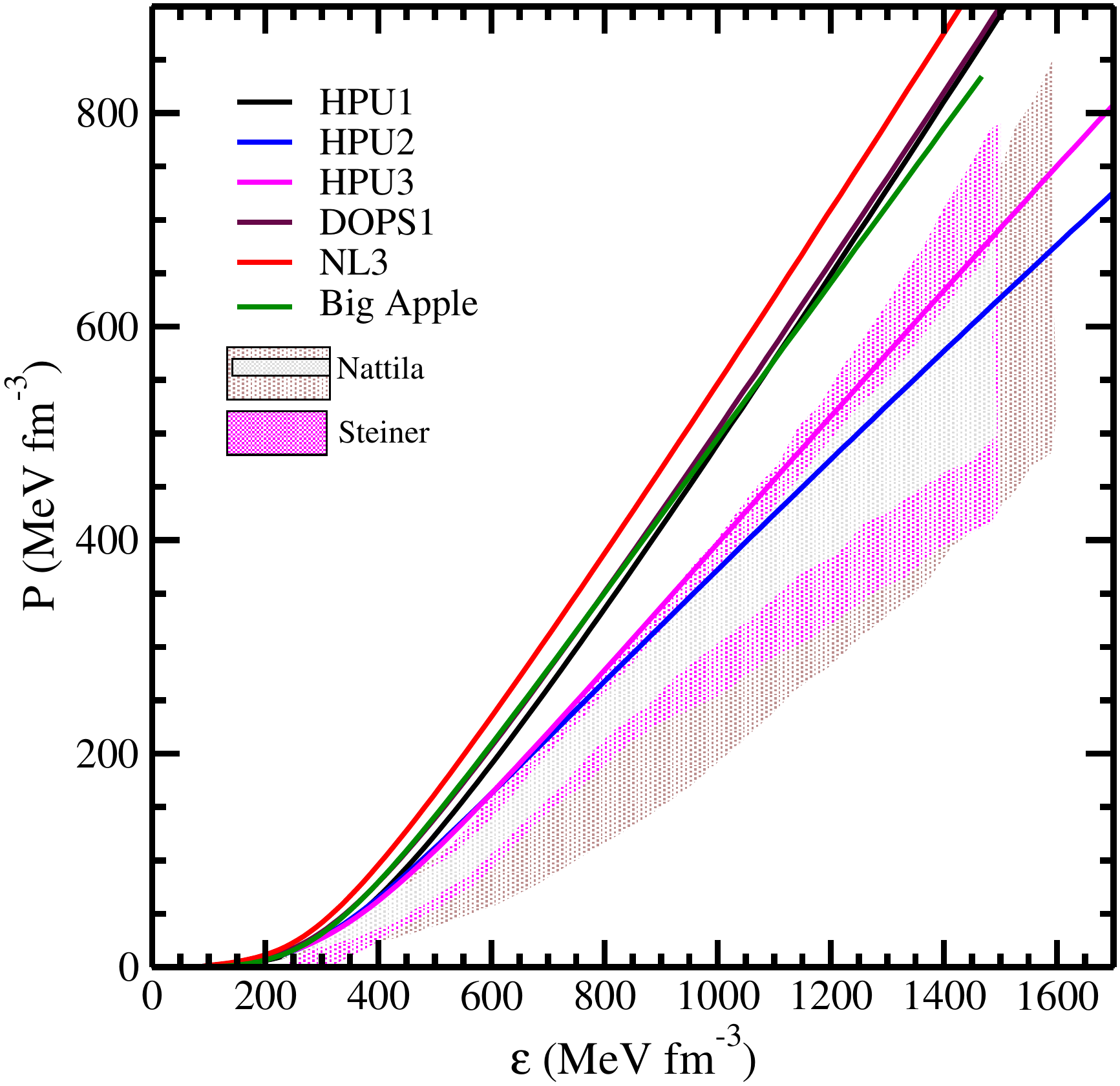}
\caption{\label{fig:eos} (color online) The range of EoSs i.e. variation of Pressure as a function of Energy Density  with  different models considered in the present work. The Shaded region (magenta) represents the observational constraints taken from reference \cite{Steiner2010} and the regions (brown and grey) represent the EoS of cold dense matter with 95\% confidence limit  \cite{Nattila2016}.}
\end{figure*}

 In Fig. \ref{fig:eos}, we display  EoS i.e. pressure as a function of  energy density  for the $\beta$ equilibrated nucleonic matter for HPUs  parametrizations. Simliar results are also shown for DOPS1, NL3 and Big Apple parameter sets. The Shaded region (magenta) represents the observational constraints at $r_{ph}$=R with the 2$\sigma$ uncertainty   \cite{Steiner2010}. Here $r_{ph}$ and R  are the photospheric and neutron star radius respectively and the regions (brown and grey) represent the EoS of cold dense matter with 95\% confidence limit reported in Ref. \cite{Nattila2016}. It can be observed that the EoSs computed with HPU2 and HPU3 parametrizations are relatively softer and lie in the allowed regions that represent the  observational constraints on  EOSs reported in  \cite{Steiner2010, Nattila2016}. The EoSs obtained by HPU1, DOPS1, NL3 and Big Apple models are showing stiff behavior and are ruled out by the shaded regions shown in Fig. \ref{fig:eos}. The stiff behavior of EOSs for these parametrizations may be attributed to the fact the $\omega $ meson self-coupling term $\zeta$ is either not  included  (HPU1, DOPS1 and NL3) or has   a very small value  (Big Apple). This is in accordance with the understanding that the  coupling term  $\zeta$ is responsible for softening of EoS at high densities \cite{Raj2006, Muller1996, Pradhan2023}. Amongst HPUs models, the EoS for HPU2 parametrization is found to be relatively softer at higher densities (Fig. \ref{fig:eos}), might be attributed to  value of coupling $\zeta$  which is somewhat larger as compared to other parameterizations.
 The mass and radius of a neutron star are obtained by solving the Tolman-Oppenheimer-Volkoff (TOV) equations \cite{Oppenheimer1939,Tolman1939} given as:

\begin{equation}
\label{eq:tov}
\frac{dP(r)}{dr} = -\frac{\{\epsilon(r)+P(r)\}\{4\pi r^3 P(r)+m(r)\}}{r^2(1-2m(r)/r)}
\end{equation}
\begin{equation}
\label{eq:nr31}
\frac{dm}{dr}=4\pi r^2\epsilon(r),
\end{equation}
\begin{equation}
m(r)= 4\pi\int_0^{r}dr r^2 \epsilon(r) 
\end{equation}
here, $P(r)$  is the pressure at radial distance $r$ and $m(r)$  is the  mass of the neutron star  enclosed in the sphere of radius $r$. The radius of canonical neutron star is more  sensitive to EoS of crust region than those of maximum mass configurations. For lower densities, we employed  Baym-Pethick-Sutherland (BPS) \cite{Baym1971} matching on to the model EoS at $\rho$ = 0.5$\rho_{0}$  and going down to 6.0$\times$ 10$^{-12}$ fm$^{-3}$. At densities larger than 0.5$\rho_{0}$ we use model EoS obtained by nucleonic and leptonic contributions.
\begin{figure*}
\hspace{-0.3cm}\includegraphics[trim=0 0 0 0,clip,scale=0.55]{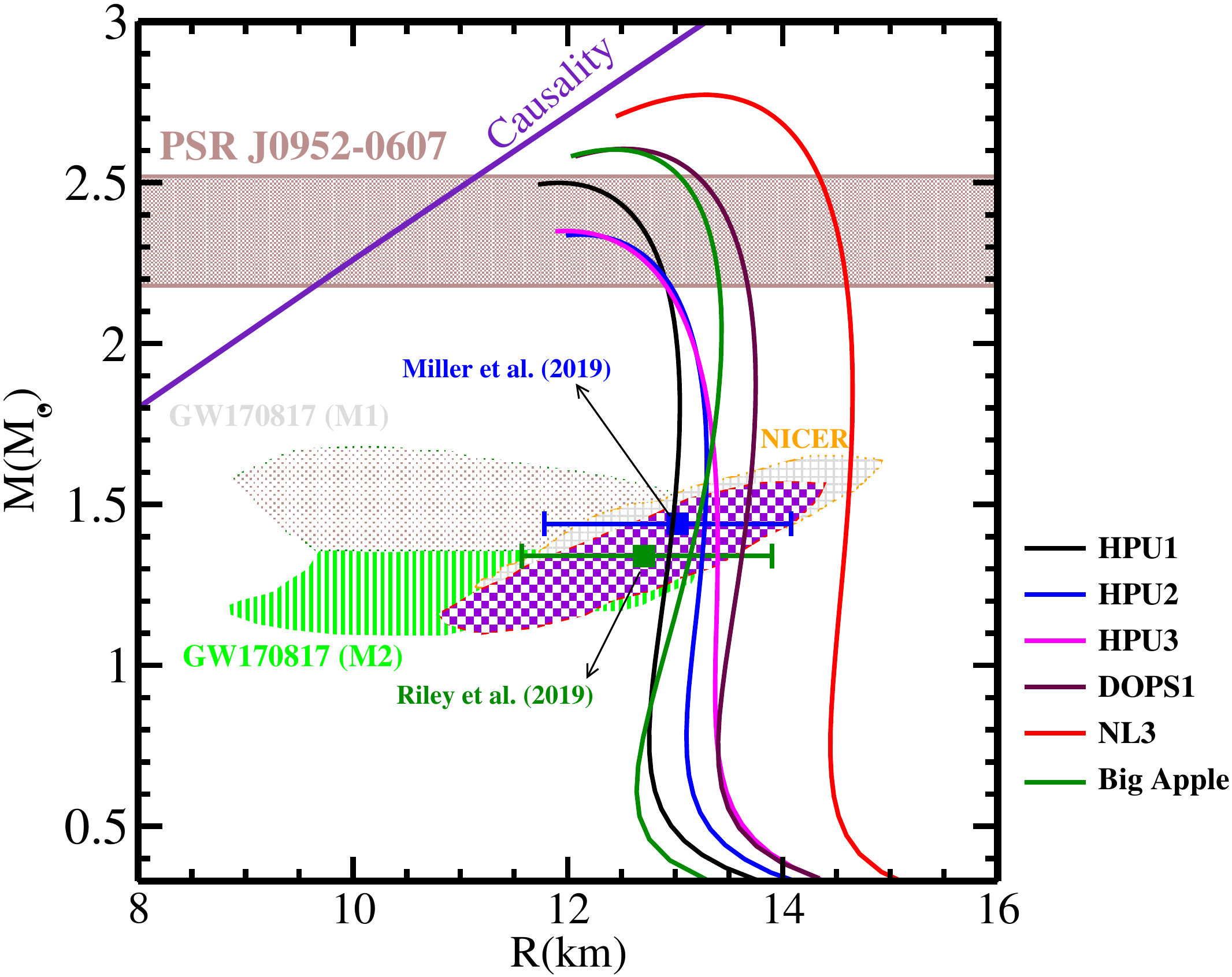}
\caption{\label{fig:mr} (color online) Neutron star mass-radius relation for various HPUs parametrizations. Horizontal bands correspond to mass M =2.35$\pm$0.17 M$_{\odot}$ of PSR J0952-0607 \cite{Romani2022}. The mass-radius estimates of the two companion neutron star in the merger event GW170817 \cite{Abbott2018} are shown by shaded regions labeled  with GW170817 M1 (M2) along with the constraints from NICER observations \cite{Miller2021,Riley2021}. The results are shown for DOPS1,  NL3, and Big Apple parameter sets.}
\end{figure*}
\begin{table*}
 \caption{\label{tab:table3}The properties of nonrotating neutron stars  obtained for the various parametrizations considered in the present work. $M_{\rm max}$ and $ R_{\rm max}$ denote the Maximum Gravitational mass and corresponding radius, respectively. The  values for $R_{1.4}$ and $\Lambda_{1.4}$ denote radius and  dimensionless tidal deformability at $1.4M_\odot$ and $R_{2.08}$ denotes radius  at $2.08 M_\odot$}
 \begin{tabular}{cccccc}
\hline
\bf{EOS}& \bf{M}& \bf{R$_{max}$ } & \bf{R$_{1.4}$}&\bf {R$_{2.08}$}&\bf{$\Lambda_{1.4}$}\\
 & (M$_{\odot}$) &(km) &(km) & (km)& \\
 \hline
HPU1&2.50&11.93&12.96&12.98&610.7\\
HPU2&2.34&12.06&13.26&13.09&699.8\\
HPU3&2.35&11.97&13.39&13.07&719.0\\
DOPS1&2.61&12.33&13.64&13.71&770.0\\
NL3&2.77&12.93&14.59&14.62&1241.6\\
Big Apple&2.60&12.25&13.18&13.42&715.51\\
\hline
\end{tabular}
\end{table*}
In Fig. \ref{fig:mr}, we present the  results for the gravitational mass of  static  neutron  stars and its radius  for HPU1, HPU2 and HPU3  parametrizations. Similar results calculated for other parameter sets are also displayed. The horizontal bands correspond to mass M =2.35$\pm$ 0.17 M$_{\odot}$ of PSR J0952-0607 \cite{Romani2022}. The mass-radius estimates of the two companion neutron star in the merger event GW170817 \cite{Abbott2018} are shown by shaded regions labeled  with GW170817 M1 (M2). The shaded regions depicting NICER observations \cite{Miller2021,Riley2021} are also shown. It is observed that the maximum gravitational mass of the static  neutron star for HPU1, HPU2 and HPU3 parameter set lie in the range  2.34 - 2.50 M$\odot$  which is in good agreement with the mass constraints reported for heaviest neutron star M$_{max}$ =2.35$\pm$ 0.17 M$_{\odot}$ for black widow pulsar  PSR J0952-0607 \cite{Romani2022}. The HPUs parametrizations also satisfy the mass-radius estimates of the two companion neutron stars as inferred in  the merger event GW170817 \cite{Abbott2018}  as shown by shaded regions labeled  with GW170817 M1 (M2). The neutron star mass-radius computed by using HPUs parametrizations are in good agreement with the NICER measurements \cite{Miller2021,Riley2021}. The radius $R_{1.4}$ corresponding to 1.4 M$_{\odot}$ neutron star lie in the range 12.96 - 13.39 Km for HPUs parametrizations and is  well consistent with the inferences on  the radius constraints from NICER \cite{Annala2018,Riley2021,Miller2021}. The radius of 2.08 M $_{\odot}$ neutron star lies in the range 12.98 -13.09 Km for HPUs parameter sets  and is in good agreement with value predicted $R_{2.08}= 13.7^{+2.6}_{-1.5}$ Km from Miller et al. \cite{Miller2021} and $R_{2.08}= 12.39^{+1.30}_{-0.98}$ from Riley et al. \cite{Riley2021}.\\
The tidal deformability ($\Lambda$) rendered by the companion stars on each other in a binary system can provide remarkable pieces of information on the EOS of neutron stars \cite{Hinderer2008,Hinderer2010}.
The tidal influences of its companion in BNS system will deform neutron stars in the binary system and, the resulting change in the gravitational potential modifies the BNS orbital motion and its corresponding gravitational wave (GW) signal. This effect on GW phasing can be parameterized by the dimensionless tidal deformability parameter,
$\Lambda_i = \lambda_i/M_i^5,$ i = 1, 2.  For each neutron star, its quadrupole moment ${\cal{Q}}_{j,k}$ must be related to the tidal field ${\cal{E}}_{j,k}$ caused by its companion
as, ${\cal{Q}}_{j,k} = -\lambda {\cal {E}}_{j,k}$, where, $j$ and $k$ are spatial tensor indices.
The dimensionless tidal deformability
parameter  $\Lambda$  of a static, spherically symmetric compact star
depends on the neutron star compactness parameter C and a dimensionless quadrupole Love number k$_{2}$ as, $\Lambda = \frac{2}{3} k_2 C^{-5}$. The $\Lambda$ critically parameterizes 
the deformation of neutron stars under the given tidal field, therefore it should depend on the EOS of nuclear dense matter. 
To measure the Love number $k_{2}$ along with the evaluation of the TOV  
equations we have to compute $y_{2} = y(R)$ with 
initial boundary condition y(0) = 2 from the first-order
differential equation   \cite{Hinderer2008,Hinderer2009,Hinderer2010,Damour2010} simultaneously,
 \begin{eqnarray}
 \label{y}
   y^{\prime}&=&\frac{1}{r}[-r^2Q-ye^{\lambda}\{1+4\pi Gr^2(P-{\cal{E}})\}-y^{2}], \\
  Q &\equiv & 4\pi Ge^{\lambda}(5{\cal{E}}+9P+\frac{{\cal{E}}+P}{c_{s}^2})
 -6\frac{e^{\lambda}}{r^2}-\nu^{\prime^2}\\
   e^{\lambda} &\equiv& (1-\frac{2 G m}{r})^{-1}\\
 \nu^{\prime}&\equiv&  2G e^{\lambda} 
       (\frac{m+4 \pi P r^3}{r^2}).
      \end{eqnarray} 
First, we get the solutions of Eq.(\ref{y}) with boundary condition, y$_{2}$ = y(R),
then the electric tidal Love
number k$_{2}$ is calculated from the expression as,
 \begin{eqnarray}
 k_{2}&=& \frac{8}{5}C^{5}(1-2C)^{2}[2C(y_{2}-1)-y_{2}+2]\{2C(4(y_{2}+1)C^4\nonumber\\
 &+&(6y_{2}-4)C^{3}
 +(26-22y_{2})C^2\nonumber\\
 &+&3(5y_{2}-8)C-3y_{2}+6)\nonumber\\
 &-&3(1-2C)^2(2C(y{_2}-1)-y_{2}+2) \log(\frac{1}{1-2C})\}^{-1}.\nonumber\\
 \end{eqnarray}
 The value of $\Lambda_{1.4}$ obtained for canonical mass with HPUs  parameter sets lie in the range 610.7 - 719.0 which satisfies the value obtained from the GW170817 event \cite{Reed2021,Abbott2017,Chen2021} for the EoS of dense nuclear matter. \\
Further, it is noted that  the our analysis of  tidal deformability  ($\Lambda_{1.4}$) lies  within  the constraints ($\Lambda_{1.4} \leq $ 800) for GW170817 event  \cite{Abbott2017}, $\Lambda_{1.4} = 575^{+262}_{-232}$ \cite{Biswas2021} using the Miller et al \cite{Miller2021} posteriors  and $\Lambda_{1.4} = 457^{+219}_{-256}$ \cite{Legred2021} using the Riley et al \cite{Riley2021}. The value of $\Lambda_{1.4}$ obtained for HPU1 model is very close to the upper limit  with revised   limit $\Lambda_{1.4} \leq  580$ within $1\sigma$ uncertainty  \cite{Abbott2018}. The precise measurement of  tidal deformability  can constrain the neutron star radius in narrow bounds. Indeed it is  believed that no terrestrial experiment can reliably constrain the EOS of neutron star \cite{Reed2021}.
In Table \ref{tab:table3}, we summarized the results for the various properties of neutron  stars obtained with HPU1, HPU2 and HPU3  parametrizations. Results obtained with other parameter sets are also shown for comparison.
\subsection{Correlations amongst  nuclear matter observable and model parameters}
\begin{figure*}
\centering
\includegraphics[trim=0 0 0 0,clip,scale=0.55]{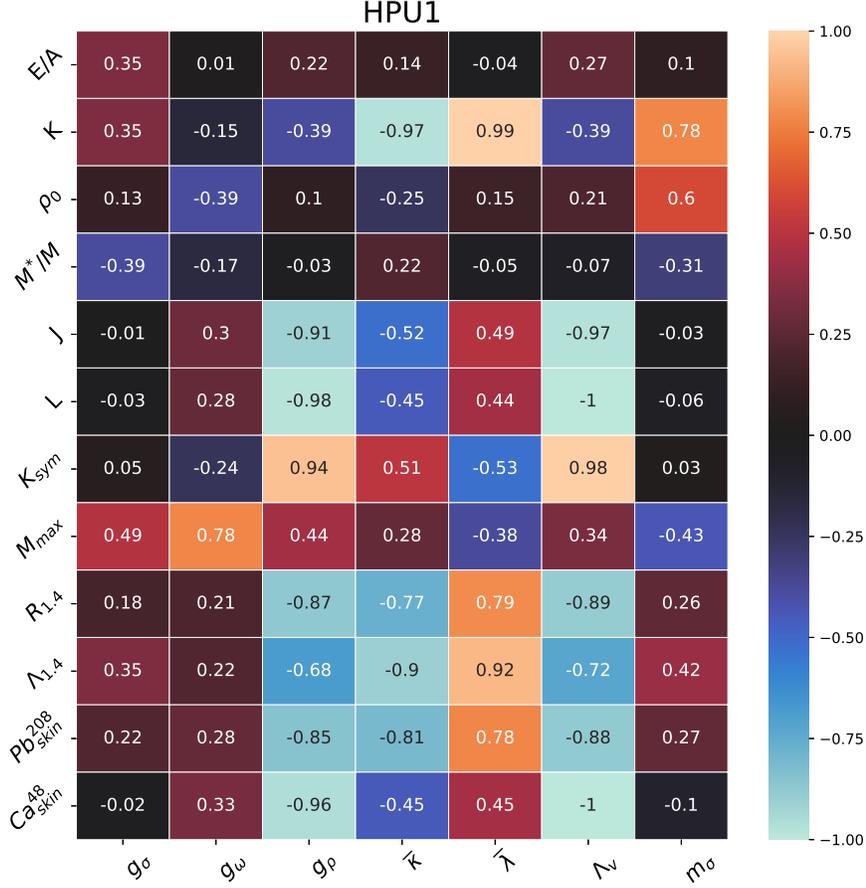}
\caption{\label{prop_hpu1} (Color online) Correlation coefficients amongst neutron star properties  as well as the  bulk properties  of nuclear matter at the saturation density and model parameters for HPU1 parametrization.}
\end{figure*}

\begin{figure*}
\centering
\includegraphics[trim=0 0 0 0,clip,scale=0.55]{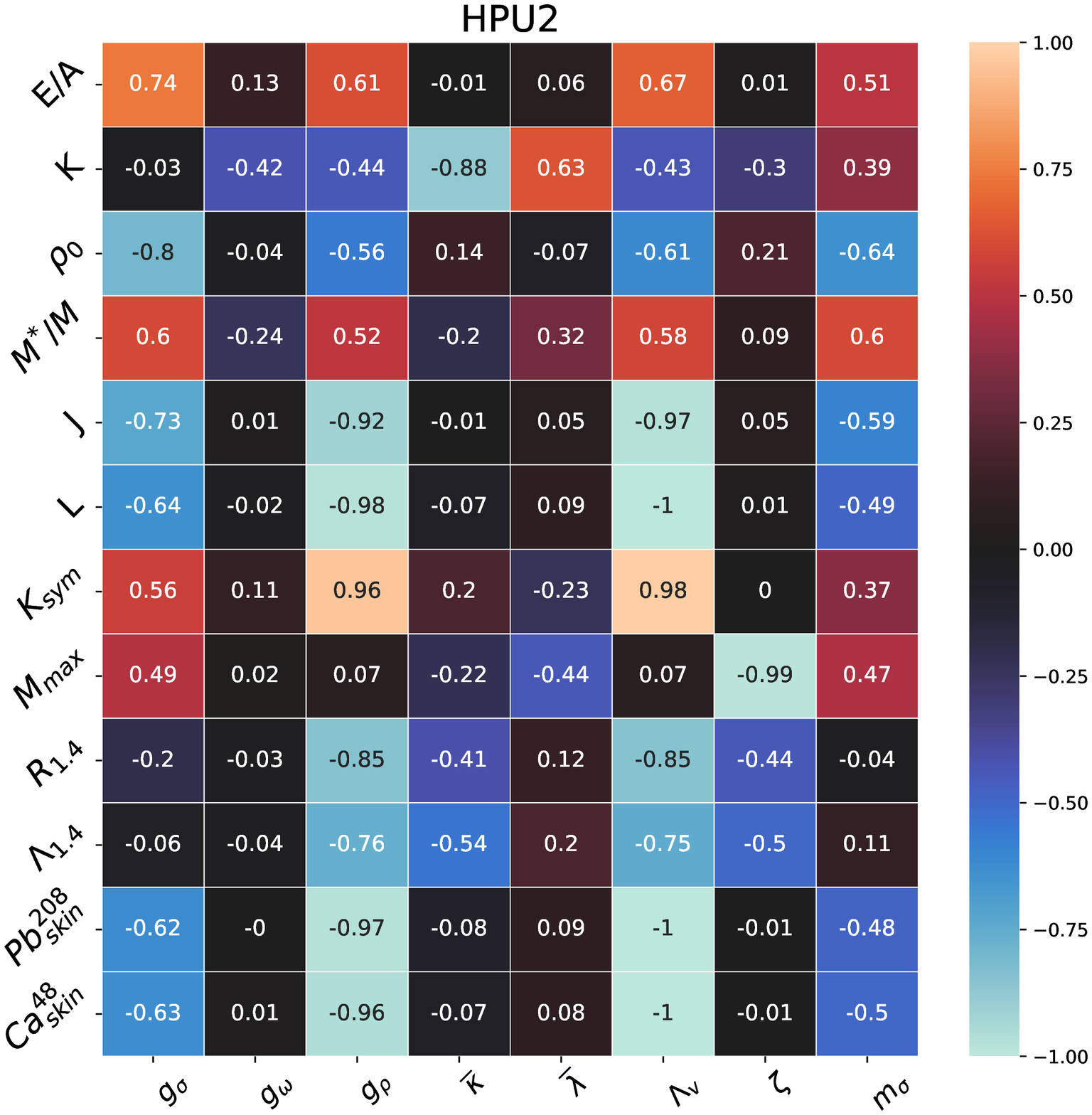}
\caption{\label{prop_hpu2} (Color online) Same as Fig. \ref{prop_hpu1}, but for HPU2 paramerization.}
\end{figure*}

\begin{figure*}
\centering
\includegraphics[trim=0 0 0 0,clip,scale=0.55]{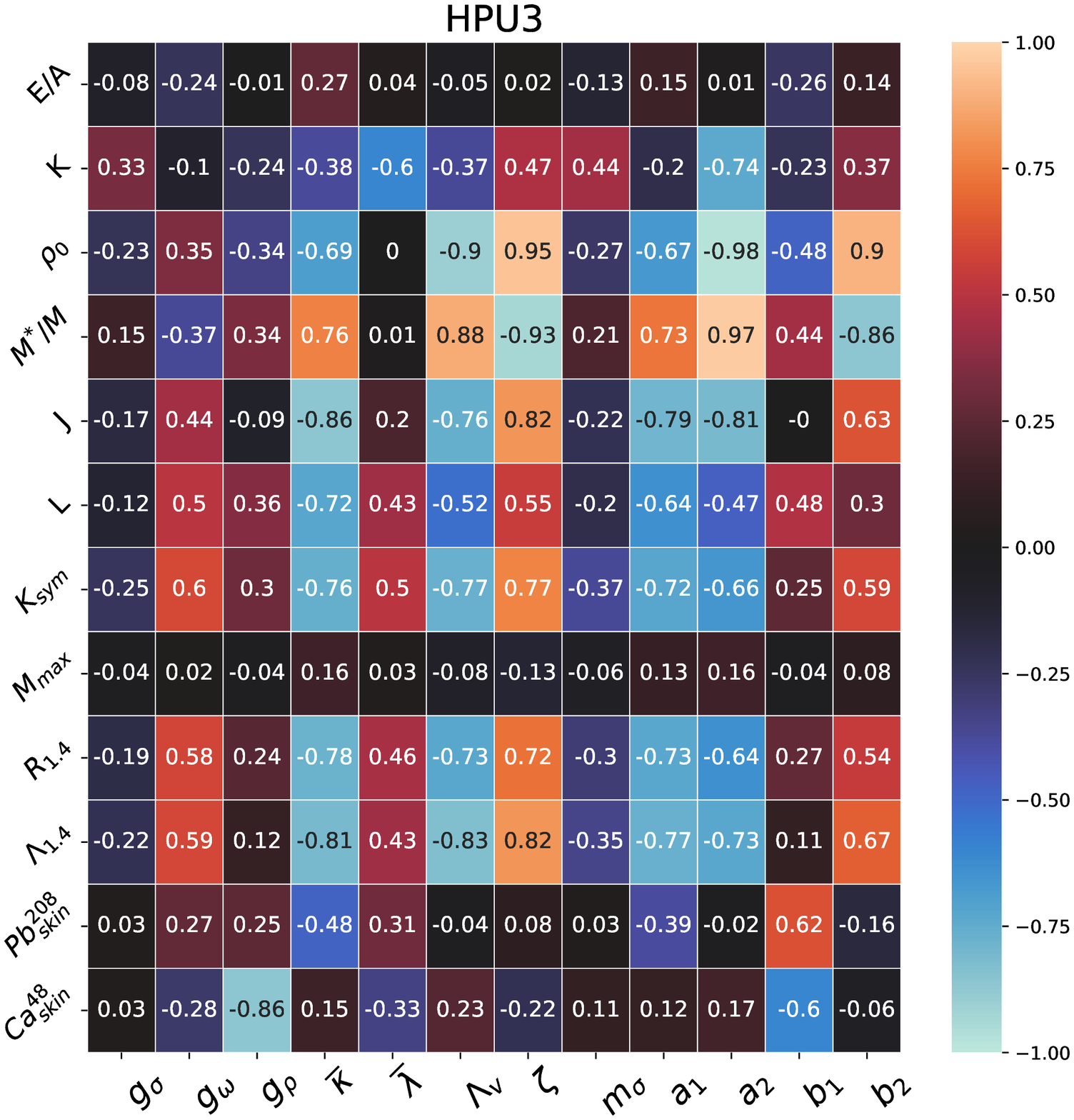}
\caption{\label{prop_hpu3} (Color online) Same as  Fig. \ref{prop_hpu1}, but for HPU3 paramerization.}
\end{figure*}

In this section, we discuss the  correlations  between the bulk  nuclear matter and neutron star observables and model parameters. In Fig. \ref{prop_hpu1} to Fig. \ref{prop_hpu3}, we display the correlations of  bulk nuclear matter properties at saturation density and neutron star observables with the model parameters for HPU's parametrizations.

For the HPU1 model, the isoscalar nuclear matter properties  K, show strong correlations with isoscalar parameters $\overline{\kappa}$ and $\overline{\lambda}$. It can also be observed from Fig.\ref{prop_hpu1} that the  J, L and  $K_{sym}$  can be very well constrained by the coupling parameter $g_{\rho}$ and $\Lambda_{v}$ as depicted by their correlations. The neutron star observables  $R_{1.4}$ and $\Lambda_{1.4}$ also show strong dependence on  couplings $\Lambda_{v}$ and $g_{\rho}$.
The neutron skin thickness of $^{208}$Pb and $^{48}$Ca also shows strong  correlation with $\Lambda_{v}$ and $g_{\rho}$ for HPU1 and HPU2 models.
For the HPU2 model, the isoscalar nuclear matter property  E/A, show good correlations with isoscalar parameters $g_{\sigma}$ and K is strongly correlated with $\overline{\kappa}$. It is evident  from the Fig. \ref{prop_hpu2} that the  isovector  properties like J, L, $K_{sym}$ and neutron star observables $R_{1.4}$ and $\Lambda_{1.4}$ have strong dependence on $g_{\rho}$ and $\Lambda_{v}$ as suggested by their correlations. A strong negative correlation is observed for neutron star maximum mass ($M_{max}$) with coupling $\zeta$ which is well consistent with the findings reported in Ref. \cite{Muller1996,Raj2006, Pradhan2023} which indicates that  the value of $\zeta$ is either zero or very small for supporting the hypermassive neutron star.
For HPU3 model, the isovector  properties like J, L, $K_{sym}$  show a weak correlation with coupling $g_{\rho}$. J is found to be  strongly correlated with $a_{1}$ and $a_{2}$ and shows moderate correlation with b$_{2}$. The $K_{sym}$ also show good dependence on cross interaction terms $a_{1}$, $a_{2}$ and $b_{2}$. The  neutron star observables like $R_{1.4}$ and $\Lambda_{1.4}$ show a good correlations with couplings $a_{1}$, $a_{2}$ and $b_{2}$ and weak correlation with $g_{\rho}$. It may be noticed from Fig. \ref{prop_hpu3} that the observables like J, L, $K_{sym}$, $R_{1.4}$ and $\Lambda_{1.4}$  show moderate to good correlations with cross coupling terms incorporated in HPU3 model. The neutron skin thickness of $^{208}$Pb and $^{48}$Ca also shows good  dependence on coupling $b_{1}$.
\section{Summary}\label{summary}
Three new parametrizations namely HPU1, HPU2 and HPU3 for the relativistic mean field model have been generated in the light of heaviest observed neutron star for the black widow pulsar PSR J092-0607, astrophysical constraints and naturalness behavior of the coupling parameters as demanded by effective field theory in addition to those usually employed, like, binding energy, charge radii  for finite nuclei  and empirical data on the nuclear matter at the saturation density. We have taken  different combinations of  non-linear, self and cross interaction terms  between $\sigma$, $\omega$, and $\rho$-meson up to the quartic order in the Lagrangian of RMF model.
The newly generated  HPUs parametrizations of the RMF model can accommodate  the properties of NSs within the astrophysical observations without compromising the finite nuclei and bulk nuclear matter  properties.
In the HPU1 parametrization, self interaction terms ($\overline {\kappa}, \overline{\lambda}$) of $\sigma$ meson and cross interaction term ($\Lambda_{v}$) of $\omega$-$\rho$ mesons, in addition to the exchange interactions of baryons with $\sigma$, $\omega$ and $\rho$ mesons are taken in the Lagrangian. The $\omega$ meson self-interaction  term  $\zeta$ is not included in order to maintain compatibility with  heaviest observed  neutron star mass  M$_{max}$= 2.35 $\pm$0.17 M$_{\odot}$ for black widow pulsar PSR J0952-0607.
 For the  HPU2 parametrization,  we  also incorporate the $\omega$ meson self coupling parameter $\zeta$ in addition to the coupling terms considered in HPU1 model. In HPU3 parametrization we include all possible self and cross-couplings among isoscalar-scalar $\sigma$, isoscalar-vector $\omega_{\mu}$ and isovector-vector $\rho_{\mu}$ meson fields up to the quartic order so that parameter set generated satisfy the  mass constrains of PSR J0952-0607. The inclusion of these possible self and cross-interactions terms are important to accommodate the naturalness behavior of parameters as imposed by effective field theory \cite{Furnstahl1997, Raj2006}.
All HPUs parametrizations  are  obtained such that it reproduces the ground state properties of the finite nuclei,  bulk properties of nuclear matter  and also satisfy the constraints of mass from the heaviest observed neutron star from the black widow pulsar \cite{Romani2022}. We notice that the nonlinear $\omega$ meson  self-coupling $\zeta$, is either very small (HPU2, HPU3) or zero for HPU1 interactions for supporting hypermassive neutron star.
 The root mean square errors in the total binding energies  for finite nuclei included in our fit for HPUs parametrizations are 3.06, 1.81 and 2.35 MeV respectively. The root mean square errors in the charge rms radii for the nuclei included in our fit are 0.050, 0.016 and 0.017 fm respectively. The Bulk nuclear matter properties obtained  are well consistent with the current empirical data \cite{Reed2021,Piekarewicz2014, Li2013}.
 The maximum gravitational mass of the neutron star for HPUs parameter sets lie in the range  2.34 - 2.50 M$\odot$ and is in accordance with the heaviest observed neutron star from the black widow pulsar PSR J0952-0607. The radius ($R_{1.4}$) of the neutron star lie in the range 12.96 - 13.39 km respectively and is in good agreement with the results reported in \cite{Miller2019,Riley2019, Miller2021, Riley2021,Annala2018}. The value of $\Lambda_{1.4}$ obtained for the HPUs parametrizations lie in the range 610.7 - 719.0 and  also satisfies the constraints for GW170817  event \cite{Abbott2017} and  reported in Refs. \cite{Chen2021,Reed2021, Biswas2021, Legred2021}.\\
 The parametrizations generated in light of PSR J0952-0607  demands  stiff EoSs leading to  the relatively  larger value of  $\Lambda_{1.4}$. To satisfy    the revised limit of tidal deformability $\Lambda_{1.4}\leq 580$  \cite{Abbott2018}, the softening of EoS at intermediate densities, together with the subsequent stiffening at high-density regions is required to support massive neutron stars that may be indicative of a phase transition in the steller core \cite{Fattoyev2018}.

\begin{acknowledgments}
Author(s) are thankful to Himachal Pradesh University for providing the computational facility. B.K.A. acknowledges partial support from the SERB, Department of science and technology, Govt. of India with grant numbers SIR/2022/000566 and CRG/2021/000101 respectively. S.K. is highly thankful to CSIR-UGC (Govt. of India) for providing financial assistance (NTA/ 211610029883 dated
19/04/2022) under Junior/Senior Research Fellowship scheme.

\end{acknowledgments}
  \bibliography{Ref}
  \bibliographystyle{unsrt}
\end{document}